# Model Checking of Boolean Process Models

Christoph Schneider[1], Joachim Wehler[2]

[1]ckes@gmx.net,
[2]joachim.wehler@gmx.net, Ludwig-Maximilians-Universität München, Cirquent GmbH

**Abstract**. In the field of Business Process Management formal models for the control flow of business processes have been designed since more than 15 years. Which methods are best suited to verify the bulk of these models?

The first step is to select a formal language which fixes the semantics of the models. We adopt the language of Boolean systems as reference language for Boolean process models. Boolean systems form a simple subclass of coloured Petri nets. Their characteristics are low tokens to model explicitly states with a subsequent skipping of activations and arbitrary logical rules of type AND, XOR, OR etc. to model the split and join of the control flow.

We apply model checking as a verification method for the safeness and liveness of Boolean systems. Model checking of Boolean systems uses the elementary theory of propositional logic, no modal operators are needed. Our verification builds on a finite complete prefix of a certain T-system attached to the Boolean system. It splits the processes of the Boolean system into a finite set of base processes of bounded length. Their behaviour translates to formulas from propositional logic. Our verification task consists in checking the satisfiability of these formulas.

In addition we have implemented our model checking algorithm as a java program. The time needed to verify a given Boolean system depends critically on the number of initial tokens. Because the algorithm has to solve certain SAT-problems, polynomial complexity cannot be expected. The paper closes with the model checking of some Boolean process models which have been designed as Event-driven Process Chains.

**Keywords**: *Base process, Boolean system, EPC, model checking, SAT-problem, verification.*

## 1    Introduction

In the field of Business Process Management during the last two decades several languages have emerged which are recommended for the modelling and execution of business processes. Examples are the languages EPC, BPEL, BPMN, YAWL or several components of UML [KNS1992, AND2003, OMG2009, AH2005, BRJ2005]. While some of these languages like EPCs (Event-driven Process Chains) or UML (Unified Modeling Language) are often used in commercial projects, others stay mainly in the academic domain.

With the term *Boolean process model* we denote a model of the control flow of a process, which employs rules of propositional logic to describe the branching of the control flow. A simple example is an alternative specified by an XOR-rule or an OR-rule. All languages mentioned above have constructs to model the activities of a process. The languages support the necessary process primitives sequence, iteration, alternative and parallelism. Some languages can even more, they are able to model distributed process states. These languages are used therefore to design Boolean process models.

The languages enjoy different degrees of formalization. In general their syntax is well-defined but often the semantics is ambiguous or lacks completeness. Therefore several authors have undertaken the effort to translate these languages to a reference language with a well-defined semantics. In most cases Petri nets or transitions systems are used as reference language.

Petri nets have been invented at around 1960. They had a formal semantics right from the beginning. The language of Petri nets does not only support the design of a static process model. Due to their token concept Petri nets are capable to simulate also the temporal development of a process, its runs. The language has enough expressive power to serve as a reference language for EPCs, BPEL, BPMN and the process languages of UML. The language YAWL (Yet Another Workflow Language), based on Petri nets too, extends ordinary Petri nets by constructs to deal with process patterns involving multiple instances, advanced synchronisation patterns, and cancellation patterns, see sect. 4.3 in [AH2005].

Apparently, the correct definition of the control flow is a necessary requirement for the subsequent execution of the model, e.g. by a workflow engine. And before extending the model by input or output data or even by the data flow, the correctness of the control flow has to be established. This paper does not deal with the data flow, it focuses on the control flow.

As soon as a process modelling language has got a well-defined semantics one can ask which formal methods are suitable to verify process models in this language. Petri net theory allows to verify models from certain subclasses of Petri nets in an efficient way. A paradigm is the mature theory of free-choice systems [DE1995].

Considered from a theoretical point of view, each ordinary Petri net can be verified by methods of Girard's linear logic. Linear logic represents in a direct way the firing rule of Petri nets: Entailment in linear logic has a "resource consuming character". Each derivation of a given formula consumes the binding of the variables of the premise and allocates a binding of the conclusion. This is in contrast to propositional logic where a logical derivation does not remove the binding from the variables of its premise. But the general purpose character of linear logic is also a handicap when the method is applied to Petri nets arising from commercial applications. We do not know of a tool which verifies such Petri nets with the reduction rules of linear logic in an efficient way.

Different methods have been developed for the verification of Boolean process models:

A large verification project has been directed by van der Aalst and his co-workers [DVV2006]. The authors checked about 10.000 EPCs from the reference model of the system SAP R/3, Version 4.6. Each EPC model was pre-processed by a reduction method. The reduced EPC was then transformed into a workflow net, a certain type of ordinary Petri net. These workflow nets were automatically checked by tools from the ProM framework for the behavioural properties soundness and relaxed soundness. In a related paper [MMN2006] the authors inform the reader that at least 5.6% of the 10.000 EPCs under consideration are faulty. Interestingly, the authors from [DVV2006] created a separate EPC to formalize their verification process of EPCs. We will take a test of this EPC in Chapter 5.

In a later paper [MA2008] the authors present additional reduction rules for EPCs. They claim: EPCs, which according to these rules can be reduced to EPCs with only XOR-connectors, are sound. The reduction is supported by a tool named xoEPC. The remaining irreducible EPCs have to be checked manually. The authors emphasize the necessity to discuss the intermediate EPCs with the modeller of the EPC during the reduction process, in order to ensure the intended semantics of the process model. We will take a test of the running example from [MA2008] in Chapter 5.

The authors of [FFJ2009] investigated more than 700 business models from industrial applications. They checked them for safeness and absence of deadlock. The process models were designed in a proprietary language combining UML activity diagrams with BPMN. The authors compare three different methods of verification: The first two methods translate the process models from the proprietary language to ordinary Petri nets. The first method then applies model checking on the base of Computation Tree Logic (CTL), while the second translates the ordinary Petri net to a workflow net, then applies structural reduction rules and eventually explores the state space. The third method translates the original process model to a workflow graph, which is then decomposed into fragments with only a single entry and a single exit (SESE decomposition). Eventually the method ends with a combination of heuristics and state space exploration. Each of the three methods is supported by a different tool: LoLA, Woflan and IBM Websphere Business Modeler. The authors present detailed reports about the number of detected faults and the performance of their methods.

In an earlier paper [LSW1998] we have introduced *Boolean systems* as a subclass of coloured Petri nets. We have demonstrated how Boolean systems can be used as a reference language for EPCs.

The present paper continues the study of Boolean systems. It shows how to verify Boolean systems. The verification will determine whether the Boolean system is well-behaved, i.e., safe and live. And by the verification of the Boolean system also the EPC is verified.

Liveness assures that a given action can be executed again and again from every reachable state of the process. Safeness assures that each local state of the system is determined by a well-defined token. In our opinion liveness and safeness are the two minimal requirements for a correct Boolean process model. They presuppose the Boolean system to be strongly connected.

Safeness of a Boolean system follows easily from the safeness of its skeleton. The skeleton is obtained by forgetting all colours of the Boolean system. A strongly connected skeleton is a T-system. The verification of T-systems is a well-established task. Much more difficult is the question of liveness of a Boolean system. High-performance algorithms to check liveness of a problem have to circumvent the state explosion problem. The number of reachable states increases exponentially with the size of the system. Strongly connected Boolean systems arise from T-systems by adding Boolean expressions as guard formulas to specify the different firing modes of the transitions. To check liveness of a safe Boolean system we proceed as follows:

First we translate the behaviour of the Boolean system to formulas from propositional logic. Then we check the satisfiability of these formulas. What is closer related than these two procedures?

Our approach is an example of model checking. This method follows the principle to formalize "system enjoys property" as "system's semantics is model of formula" [Esp1994]. In general the formulas in question have to be taken from modal logic. However, to analyze liveness of Boolean systems it is sufficient to employ propositional logic only, which is an elementary theory. No use of any modal operator is necessary.

Model checking of a safe, strongly connected Boolean system starts with applying prefix theory to the skeleton. One obtains a finite complete prefix of its unfolding. By adding colours the prefix extends to a Boolean net. One has to consider a finite set of base markings on it and to check deadlock freeness and liveness for each of the resulting base processes.

We have implemented our model checking algorithm by a java program and tested its performance on a standard notebook with 2.53 GHz. Our implementation of the model checking algorithm shows a performance of some seconds per model with about 25 Boolean transitions and 30 places. Of course this result is not yet comparable to the time range of milliseconds reported in [FFJ2009]. At this stage the performance bottleneck is our simple SAT-solver written on the basis of the resolvent algorithm. Of course the SAT-problem is NP-complete, nevertheless the first step to enhance the performance of our model checking implementation would be to link one of the SAT-solvers from the SAT research community.

Figure 1 shows the running example of the present paper.

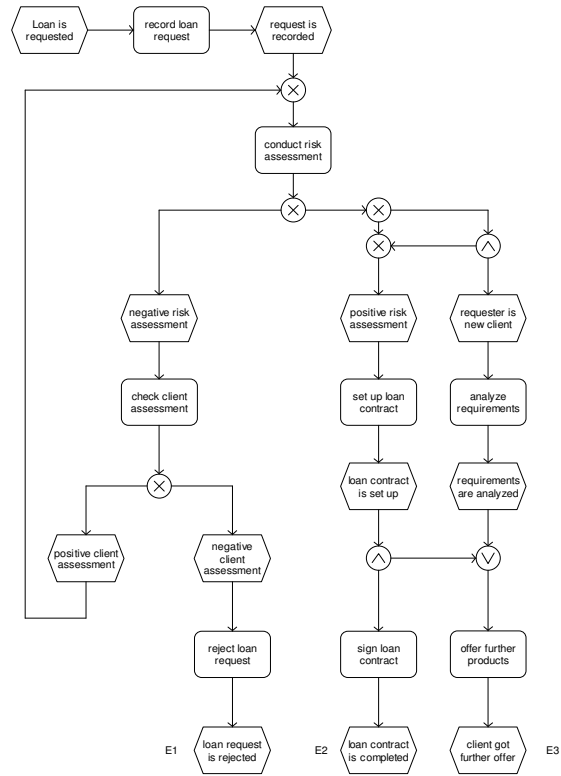

*Figure 1: EPC Loan request*

The EPC of the process "Loan request" has been taken from Fig. 1 in [MA2008] and slightly adapted. In a similar form it has been considered before in Abb. 4.31 from [Rum1999]. The process is described in [MA2008] as follows:

"The start event *loan is requested* signals the start of the process and the precondition to execute the *record loan request* function. After the post-condition *request is recorded*, the process continues with the function *conduct risk assessment* after the XOR-join connector. The subsequent XOR-split connector indicates a decision. In case of a *negative risk assessment*, the function *check client assessment* is performed. The following second XOR-split marks another decision: in case of a *negative client assessment* the process ends with a rejection of the loan request; in case of a *positive client assessment*, the *conduct risk assessment* function is executed a second time under consideration of the positive client assessment. If the risk assessment is not negative, there is another decision point to distinguish new clients and existing clients. In case of an existing client, the *set up loan contract* function is conducted. After that, the AND-split indicates that two activities have to be executed: first, the *sign loan contract* function; and second, the *offer further products* subsequent process [...]. If the client is new, the *analyze requirements* function has to be performed in addition to setting up the loan contract. The OR-join waits for both functions to be completed if necessary. If the *analyze requirements* function will not be executed in the process, it continues with the subprocess immediately [...]."

While the process starts with a unique event "Loan is requested" it ends with one or more of the three events "loan request is rejected" (E1), "loan contract is completed" (E2) and "client got further offer" (E3). E.g., not both events E1 and E2 can happen. The modeller intended either E1 or the combination of E2 and E3 as possible final events. The process comprises a loop which is executed whenever the client is assessed positively but his requested loan is considered too risky. Note the subtle logic of the connectors after the function "conduct risk assessment: Either the event "negative risk assessment" happens or the event "positive risk assessment". In the latter case, the event "requester is new client" may occur in addition.

The rest of the paper is structured as follows. Section 2 recalls some fundamental concepts from the theory of ordinary Petri nets, in particular their prefix theory. Section 3 introduces the class of Boolean systems, a subclass of coloured Petri nets. We will use Boolean systems as a reference language for Boolean process models in general and EPCs in particular. Section 4 introduces the colouring of prefixes and the base processes of a safe Boolean system. We present a model checking algorithm as the main result of our paper. We apply the results in section 5 to the verification of EPCs. The paper continues in section 6 with comparing our method to the methods above proposed for the verification of EPCs. The paper ends with an outlook to future research.

We assume that the reader is familiar with the theory of ordinary Petri nets.

## 2  Ordinary Petri nets and their processes

For the convenience of the reader and to fix the notation we recall some fundamental concepts from the theory of ordinary Petri nets, see also [DE1995].

A finite *ordinary Petri net* is a pair $(N, \mu)$: The *net* $N = (P, T, F)$ comprises a finite set $P$ of *places*, a disjoint finite set $T$ of *transitions* and a set $F \subseteq (P \times T) \cup (T \times P)$ of *directed arcs*. The function $\mu : P \longrightarrow \mathbf{N}$ is named the *initial marking* of the net. The *support* of the marking $\mu$ is the set

$$supp(\mu) := \{\, p \in P : \mu(p) > 0 \,\}$$

of all places marked at $\mu$. We will often dispense with an explicit notation for the set of places and transitions of a net and use the shorthand $x \in N$ to denote a node $x \in P \cup T$. For a node $x \in N$ we denote respectively by

$$pre(x) := {}^{\bullet}x = \{\, y \in N : (y, x) \in F \,\} \text{ and } post(x) := x^{\bullet} = \{\, y \in N : (x, y) \in F \,\}$$

the pre-set and post-set of $x$. For a subset $X \subset N$ we set

$$pre(X) := \bigcup_{x \in X} pre(x) \text{ and } post(X) := \bigcup_{x \in X} post(x).$$

All nets $N = (P, T, F)$ will be assumed *connected*, i.e. every two nodes $x, y \in N$ satisfy $(x, y) \in (F \cup F^{-1})^*$. Within the net $N$ a *path* from a node $x_{ini} \in N$ to a node $x_{fin} \in N$ is a sequence $(x_0, x_1, ..., x_n)$ with nodes $x_i \in N$, $x_0 = x_{ini}$, $x_n = x_{fin}$ and $(x_i, x_{i+1}) \in F$. The path is named *elementary path*, if $x_i \neq x_j$ for all pairs $i \neq j$. A *circuit* is a path $(x_0, x_1, ..., x_n)$ with $x_n = x_0$, it is named *elementary circuit* if the path $(x_0, x_1, ..., x_{n-1})$ is elementary. The net $N$ is *strongly connected* if for every two nodes $x_1, x_2 \in N$ a path from $x_1$ to $x_2$ and a path from $x_2$ to $x_1$ exists.

For a net $N$ the *firing rule* defines the firing of a transition: A transition $t \in T$ is *enabled* at a marking $\mu$ of $N$ iff each place from its pre-set $pre(t)$ is marked at $\mu$ with at least one token. Being enabled, $t$ may *occur* or *fire*. Firing $t$ yields a new marking $\mu'$, which results from $\mu$ by consuming one token from each pre-place of $t$ and by producing one token on each post-place of $t$; this is denoted by $\mu \xrightarrow{t} \mu'$.

A finite *occurrence sequence* from $\mu$ is a sequence $\sigma = t_1...t_k$, $k \in N$, such that

$$\mu \xrightarrow{t_1} \mu_1, ..., \mu_{k-1} \xrightarrow{t_k} \mu_k.$$

We denote by $\mu \xrightarrow{\sigma} \mu_k$ the fact, that firing $\sigma$ at the marking $\mu$ yields the marking $\mu_k$. If

$$\mu \xrightarrow{t_1} \mu_1 \xrightarrow{t_2} \mu_2 \xrightarrow{t_3} ...$$

for an infinite sequence $\sigma = t_1 \cdot t_2 \cdot t_3...$ then $\sigma$ is named an infinite occurrence sequence from $\mu$. A *reachable marking* of a Petri net $(N, \mu)$ is a marking, which results from firing a finite occurrence sequence from $\mu$. If not stated the contrary, occurrence sequences in this paper will be considered to be finite. The transitions from an occurrence sequence $\sigma = t_1...t_k$ are not necessary pair wise different. The *concatenation* of two occurrence sequences $\sigma_1$ and $\sigma_2$ is denoted by $\sigma_1 \cdot \sigma_2$.

A Petri net $(N, \mu_0)$ is *live* iff for each reachable marking $\mu$ and for each transition $t \in T$ the Petri net $(N, \mu)$ has a reachable marking which enables $t$. A Petri net is *bounded* iff there exists a natural number which bounds from above the token content of every place at every reachable marking. If the bound can be chosen equal to $1$, then the Petri net is named *safe*.

A simple class of ordinary Petri nets are marked synchronization graphs or T-systems. They are important for the present investigation because T-systems will be the skeletons of strongly connected Boolean systems introduced in Chapter 3.

### 2.1 Definition *(T-system)*

A net $N$ is a *T-net* if all places have exactly one pre-transition and exactly one post-transition, i.e.

$$|pre(p)| = 1 = |post(p)| \text{ for all places } p \in N.$$

A *T-system* is a Petri net $(N, \mu)$ with $N$ a T-net.

If the firing of a finite occurrence sequence $\mu_1 \xrightarrow{\sigma} \mu_2$ in a Petri net $(N, \mu_0)$ reproduces the original marking, i.e. if $\mu_2 = \mu_1$, then the multiset

$$Parikh(\sigma) := \sum_{t \in \sigma} t$$

formed by all transitions from $\sigma$ is a T-invariant. Each T-invariant of a connected T-net $N = (P, T, F)$ is a multiple of

$$\tau_N := \sum_{t \in T} t ,$$

the multiset of all transitions from $T$, see Prop. 2.37 and Prop. 3.16 in [De1995].

A live T-system $(N, \mu_0)$ is *cyclic*, i.e. for each reachable marking $\mu$ of $(N, \mu_0)$ the initial marking $\mu_0$ is reachable in $(N, \mu)$. This result follows at once from Theor. 3.21 in [DE1995].

A useful means to control all reachable states of a Petri net is the concept of its unfolding and the corresponding prefix theory. For the convenience of the reader we recall now some relevant definitions and properties; see also [Esp1994, EH2008].

Let $(P, T, F)$ be a net and let $x_1, x_2 \in P \cup T$. The nodes $x_1$ and $x_2$ are in *conflict*, denoted by $x_1 \# x_2$, if distinct transitions $t_1, t_2 \in T$ exist with $pre(t_1) \cap pre(t_2) \neq \emptyset$ and $(t_1, x_1), (t_2, x_2)$ belonging to the reflexive and transitive closure of $F$. For $x \in P \cup T$, $x$ is in *self-conflict* if $x \# x$.

An *occurrence net* is a net $ON = (B, E, K)$ such that

$$|pre(b)| \leq 1 \text{ for all } b \in B,$$

the (irreflexive) closure of $K$ is acyclic, no element $x \in E$ is in self-conflict and $ON$ is well-founded, i.e. for every $x \in B \cup E$, the set of elements $y \in B \cup E$ such that $(y, x)$ belongs to the transitive closure of $K$ is finite. Elements of $E$ are called *events* (German: <u>E</u>reignis), elements of $B$ *conditions* (German: <u>B</u>edingung) and $K$ is named the *causal dependency relation* (German: <u>K</u>ausalitätsbeziehung). If in addition also

$$|post(b)| \leq 1 \text{ for all } b \in B$$

then the occurrence net is called a *causal net*.

Because $ON$ is acyclic the relation $K$ is a partial order on $B \cup E$, which we denote by $\prec$. Its reflexive and transitive closure is denoted by $\preceq$. Due to the well-foundedness of $ON$ the set $min(ON)$ of minimal elements with respect to $\prec$ is non-empty for non-empty $ON$. A set $B'$ of conditions of $ON$ is a *co-set* if

$$\text{for all } b, b' \in B': not\ (b \prec b')\ and\ not\ (b' \prec b)\ and\ not\ (b \# b').$$

A maximal co-set $B'$ with respect to set inclusion is called a *cut* of $ON$.

Causal and occurrence nets are the technical means to abstract from the concept of an occurrence sequence with a well-determined order of firing its transitions to the concept of a process, which does no longer distinguish between occurrence sequences differing only by the interleaving of their transitions. A further step is the introduction of branching processes which represent in compact form a set of alternative processes. And the final step is to prove the existence of a unique maximal branching process which is named the unfolding of the original Petri net.

### 2.2    Definition *(Processes and branching processes)*

Consider a Petri net $(N, \mu)$ with $N = (P, T, F)$.

i) A *branching process* $(ON, pr)$ of $(N, \mu)$ is a discrete morphism

$$pr : ON \longrightarrow N$$

with $ON = (B, E, K)$ an occurrence net and such that $\min(ON)$ corresponds to the initial marking $\mu$, i.e.

$$\mu(p) = \left| pr^{-1}(p) \cap \min(ON) \right| \text{ for all places } p \in P.$$

A transition $t \in T$ *occurs* in the process $pr$ iff $t \in pr(T)$. A *process* $(ON, pr)$ net is a branching process if $ON$ is a causal net.

ii) On the set of all branching processes of $(N, \mu)$ the inclusion of nets defines a partial order:

$$(pr_1 : ON_1 \longrightarrow N) \subseteq (pr_2 : ON_1 \longrightarrow N) :\Leftrightarrow ON_1 \subseteq ON_2 \text{ and } pr_1 = pr_2 \mid ON_1.$$

In this case $pr_1$ is named a *prefix* of $pr_2$. A maximal branching process of $(N, \mu)$ is named an *unfolding* of $(N, \mu)$.

A branching process $pr : ON \longrightarrow N$ of $(N, \mu)$ maps each event of $ON$ onto a transition of $N$ and each condition of $ON$ onto a place of $N$. In this sense an event $e \in ON$ represent the firing of the transition $pr(e) \in N$ and a condition $b \in ON$ represents a token on place $pr(b) \in N$. Each cut $B'$ of $ON$ corresponds to the reachable marking $mark(B')$ of $(N, \mu)$, which marks each place of $p \in N$ with $\left| pr^{-1}(p) \cap B' \right|$ tokens.

Each safe Petri net has an unfolding which is uniquely determined up to isomorphism [Eng1991]. In general the unfolding is an infinite net. But the unfolding of a safe Petri net always has finite complete prefixes [McM1995]. They serve as a substitute for the unfolding, because they represent each reachable marking and the firing of each transition, which can occur in the original Petri net.

### 2.3 Definition *(Complete prefix)*

Consider the unfolding $pr : Unf \longrightarrow N$ of a Petri net $(N, \mu_0)$. A prefix

$$pr \mid ON \longrightarrow N, ON \subseteq Unf,$$

is *complete* iff

- for every reachable marking $\mu$ of $(N, \mu_0)$ a cut $c$ of $ON$ exists with

$$\mu(p) = \left| pr^{-1}(p) \cap c \right| \text{ for all places } p \in N$$

- and for every transition t of $N$, which can occur in $(N, \mu_0)$, an event $e$ of $ON$ exists with $t = pr(e)$.

It is the first condition in Definition 2.3, which will be relevant for the model checking algorithm in Chapter 4. The first condition assures that each reachable marking of the Petri net is already reachable by a subprocess of a complete prefix.

### 2.4 Example *(Finite complete prefix)*

Figure 2 shows a finite complete prefix $pr : ON \longrightarrow N$ of a live and safe T-system $(N, \mu_0)$

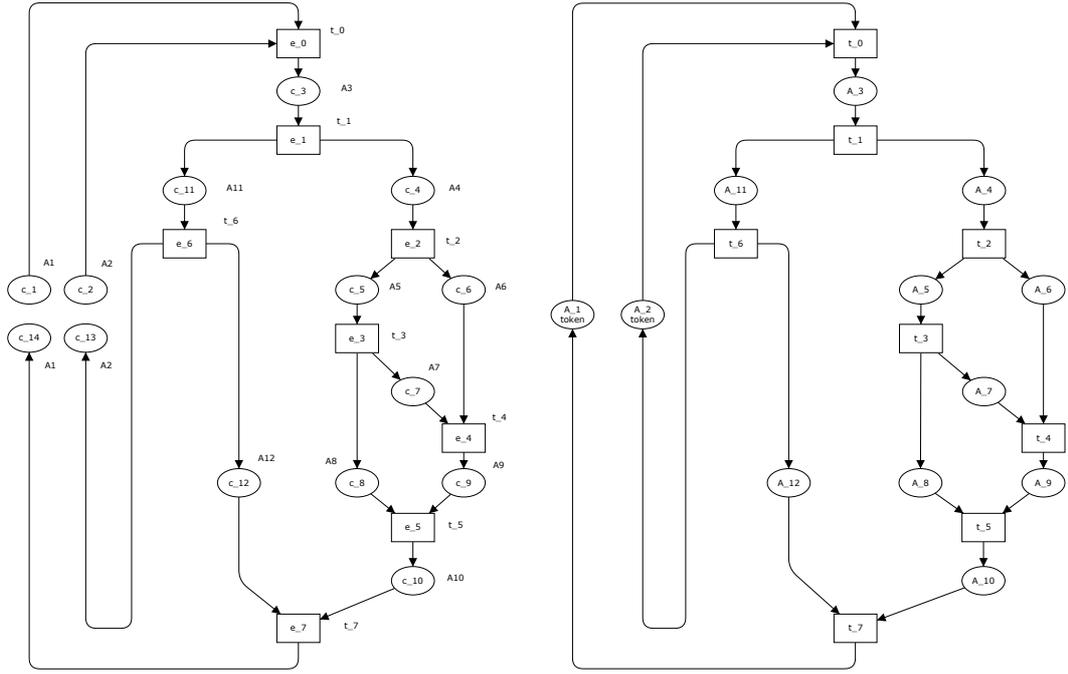

*Figure 2: Finite complete prefix of a live and safe T-system (right)*

The process is visualised as follows: Nodes of $ON$ at the left hand side as well as nodes of $(N, \mu_0)$ at the right hand side are annotated with letters inside the node. In addition each node $x \in ON$ is annotated outside the node with the name of the corresponding node $pr(x) \in N$. Each of the two cuts of $ON$

$$\min ON = \{c_1, c_2\} \text{ and } \max ON = \{c_{13}, c_{14}\}$$

represents the initial marking

$$\mu_0 = mark(\min ON) = mark(\max ON).$$

## 3 Boolean Systems

In the following we denote by $BOOLE$ the set of all formulas from propositional logic over a fixed alphabet. In particular, these formulas contain the logical connectors AND ($\wedge$), XOR ($\overset{\bullet}{\vee}$), OR ($\vee$) and NOT ($\neg$). We denote by

$$Boole = \{true, false\}$$

the two-element set of truth values. We will often use *high* or the cipher *1* as a synonym for *true* and *low* or the cipher *0* as a synonym for *false*.

A Boolean net arises from an ordinary net with unbranched places by adding

- to each transition of the ordinary net a Boolean formula as guard formula which specifies different firing modes of the transition
- and to each place of the ordinary net a second colour of low tokens.

Boolean systems are a simple class of coloured Petri nets. For the purpose of the present paper we do not need to enter into the general theory.

### 3.1 Definition *(Structure of a Boolean System and skeleton)*

i) A *Boolean net* is a tuple $BN = (N, X, G)$ comprising:

- An ordinary net $N = (P, T, F)$ with unbranched places, i.e.

$$|pre(p)| \leq 1 \text{ and } |post(p)| \leq 1 \text{ for all places } p \in P,$$

- a place annotation, which annotates each place $p \in P$ with the set *Boole*,

- an arc annotation $X : F \longrightarrow Var(BOOLE)$, which maps each arc $a \in F$ to a Boolean variable $X(a) \in Var(BOOLE)$

- and a transition annotation $G : T \longrightarrow BOOLE$, which maps each transition $t \in T$ to a formula $G_t = G_t(X_1, ..., X_n, Y_1, ..., Y_m)$, its *guard formula*. The variables $\{X_1, ..., X_n\} = \{X(p,t): p \in P \text{ and } (p,t) \in F\}$ annotate the incoming arcs of $t$ and the variables $\{Y_1, ..., Y_m\} = \{X(t,p): p \in P \text{ and } (t,p) \in F\}$ annotate the outgoing arcs of $t$. The pair $(t, G_t)$ is named a *Boolean transition*, the guard formula determines the *logical type* of the transition.

ii) A *marking* $\mu$ of $BN$ is a map

$$\mu : P \longrightarrow Boole_N$$

which maps each place $p \in P$ to a multi-set $\mu(p) \in Boole_N$ over the set *Boole*. Therefore a place $p \in P$ has at $\mu$ the *token content*

$$\mu(p) = n_1 \cdot high + n_2 \cdot low \in Boole_N$$

with non-negative integers $n_1 \in N$, the number of *high tokens*, and $n_2 \in N$, the number of *low tokens*. The set of places with non-zero token content

$$supp(\mu) := \{p \in P : \mu(p) \neq 0\}$$

is named the *support* of $\mu$.

iii) A *Boolean system* is a pair $BS = (BN, \mu)$ with a Boolean net $BN$ and a marking $\mu$ of $BN$ with at least one high token.

iv) The *skeleton* $BS^{skel}$ of a Boolean system $BS = (BN, \mu)$ with Boolean net $BN = (N, X, G)$ and ordinary net $N = (P, T, F)$ is the ordinary Petri net $BS^{skel} = (BN^{skel}, \mu^{skel})$ with $BN^{skel} := N$ and with marking

$$\mu^{skel} : P \longrightarrow N, \mu^{skel}(p) := n_1 + n_2 \text{ for } p \in P \text{ with } \mu(p) = n_1 \cdot high + n_2 \cdot low \in Boole_N.$$

Forgetting all colours induces a canonical morphism of Petri nets $skel : BS \longrightarrow BS^{skel}$.

In case a place has both an outgoing and an ingoing arc, we will always annotate both arcs with the same variable. In case a place has exactly one ingoing and exactly one outgoing arc, the arc annotation will be positioned in figures inside the place.

A transition $t \in BN$ with a unique pre-place and a unique post-place is named an *unary* transition. Besides its low binding an unary transition has a unique high binding. Transitions with either two pre-places and a unique post-place or with a unique pre-place and two post-places are named respectively *closing* or *opening binary* transitions. Without loss of generality we will often restrict to Boolean systems with only binary and unary transitions. For the purpose of verification we can even skip the unary transitions.

Definition 3.1 requires that the initial marking of a Boolean system $BS$ comprises at least one high token. Otherwise no action would take place in the process represented by $BS$.

**Note**. Readers interested in the formal definition of a Petri net morphism are referred to

[Weh2006].

## 3.2 Example *(Boolean system)*

i) Figure 3 shows the scheme of a binary Boolean transition and explains its guard formula and the resulting binding elements.

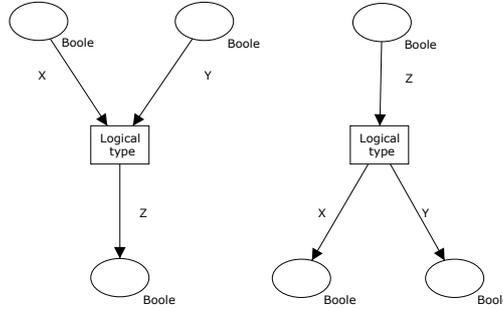

*Figure 3: Binary Boolean transitions with pre- and post-places and arc-annotations*

The column "Bindings" in Table 1 looks ahead to Definition 3.3.

| Logical type | Guard formula | Bindings for $(X,Y,Z)$ |
|---|---|---|
| AND | $(X \wedge Y \wedge Z) \vee [\neg (X \vee Y \vee Z)]$ | $(1,1,1), (0,0,0)$ |
| XOR | $\left[\left(X \stackrel{\bullet}{\vee} Y\right) \wedge Z\right] \vee [\neg (X \vee Y \vee Z)]$ | $(1,0,1), (0,1,1), (0,0,0)$ |
| AND_XOR | $[((X \wedge Y) \vee (X \wedge \neg Y)) \wedge Z] \vee [\neg (X \vee Y \vee Z)]$ | $(1,1,1), (1,0,1), (0,0,0)$ |
| XOR_AND | $[((X \wedge Y) \vee (\neg X \wedge Y)) \wedge Z] \vee [\neg (X \vee Y \vee Z)]$ | $(1,1,1), (0,1,1), (0,0,0)$ |
| OR | $[(X \vee Y) \wedge Z] \vee [\neg (X \vee Y \vee Z)]$ | $(1,0,1), (0,1,1), (1,1,1), (0,0,0)$ |

*Table 1: Guard formulas and bindings of binary Boolean transitions*

Each guard formula is valid for both the opening and the closing Boolean transition of the given logical type.

ii) Figure 4 shows a Boolean system $BS = (BN, \mu)$. The initial marking $\mu$ marks the place $A_1$ with one high token and the place $A_2$ with one low token.

The Boolean system contains an XOR-loop, a loop which is entered by an opening XOR-transition and left by a closing XOR-transition. Boolean systems also allow loops with different logical transitions like OR or AND. Note that each elementary circuit in the Boolean system from Figure 4 is marked with a single token.

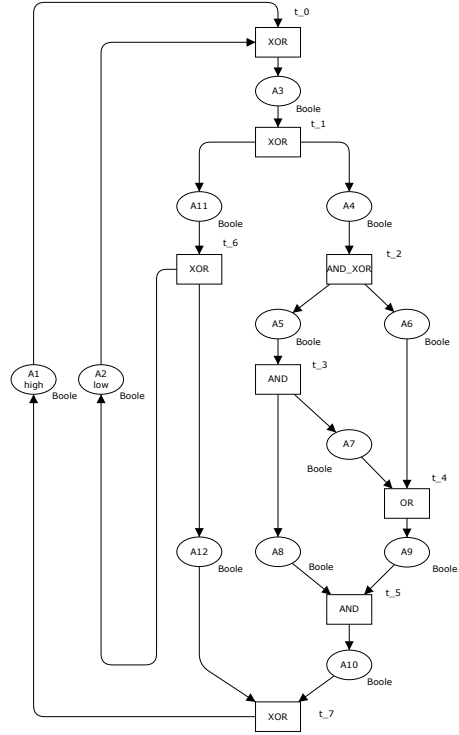

*Figure 4: Boolean system*

Different from a transition in ordinary Petri net a Boolean transition may have different firing modes. Each firing mode is named a binding element.

### 3.3 Definition *(Binding elements of a Boolean net)*

Consider a Boolean net $BN = (N, X, G)$ and a Boolean transition $(t, G_t)$ of $BN$.

i) Each binding of the variables $X_1,...,X_n, Y_1,...,Y_m$, which $G_t$ evaluates to *true*, is named a *binding* of the transition $t$ and the pair $(t,b)$ is named a *binding element*. The binding with all variables bound to *false* is named the *low binding*, all other bindings are named *high bindings*. The corresponding binding elements are named respectively *high binding element* and *low binding element*. The set of all bindings of $t$ is denoted by $B(t)$. We require that $B(t)$ contains the low binding.

ii) We require that $(t, G_t)$ is

- *faithful with respect to activation*: No high binding $(x_1,...,x_n, y_1,...,y_m) \in B(t)$ exists with

$$(x_1,...,x_n) = (0,...,0) \in Boole^n \text{ or } (y_1,...,y_m) = (0,...,0) \in Boole^m,$$

- and *fair*: For each pair $(i, j)$, $i = 1,...,n$, $j = 1,...,m$, a high binding $(x_1,...,x_n, y_1,...,y_m) \in B(t)$ exists with $x_i = y_j = 1$.

It depends on the token colours on the pre-places of the transition whether a certain binding element is enabled at a certain marking. Each binding element $(t,b)$ of $BN$ has a well-defined *enabling marking*: The marking $\mu$ of $BN$ with $supp(\mu) = pre(t)$, which enables $(t,b)$ and marks each place from $pre(t)$ with exactly one token.

### 3.4 Definition *(Firing rule of a Boolean system)*

Consider a Boolean system $BS = (BN, \mu)$ and a transition $t \in BN$ with altogether $k \in N$ pre-and post-places. A binding element

$$(t,b), \; b = (x_1,...,x_k) \in B(t) \subset Boole^k$$

is *enabled at* $\mu$, iff for each pre-place $p \in pre(t)$ with arc annotation $X(p,t) = X_i$ for an index $i \in \{1,...,k\}$ the token content

$$\mu(p) = n_1 \cdot high + n_2 \cdot low \in Boole_N$$

has coefficients respectively

$$n_1 \geq 1 \text{ if } x_i = 1 \text{ and } n_2 \geq 1 \text{ if } x_i = 0.$$

An enabled binding element $(t,b)$ may *occur*. Its *occurrence* or *firing* yields a new marking $\mu_1$: It results from $\mu$ by consuming a high token from each pre-place $p \in pre(t)$ with $x_i = 1$ and a low token from each pre-place with $x_i = 0$, and by producing at each post-place $p \in post(t)$ with arc annotation $X(t,p) = X_j$ for an index $j \in \{1,...,k\}$ a high-token if $x_j = 1$ and a low-token if $x_j = 0$. This is denoted by

$$\mu \xrightarrow{(t,b)} \mu_1.$$

By definition, each transition $t \in BN$ has a well-defined low binding $b_{low} \in B(t)$: Firing $(t, b_{low})$ consumes a low token from each pre-place of $t$ and creates a low token at each post-place of $t$. Firing the low binding is interpreted as skipping the action represented by the transition. Because the initial marking $\mu_0$ of a Boolean system $BS = (BN, \mu_0)$ contains at least one high token and because $BN$ is faithful with respect to activation, each reachable marking $\mu$ of $BS$ also contains at least one high token.

The concept of occurrence sequences which has been introduced for ordinary Petri nets in Section 2 generalizes to a Boolean system $BS = (BN, \mu_0)$: A finite *occurrence sequence from* $\mu_0$ is a sequence $\sigma = (t_1, b_1)...(t_k, b_k)$, $k \in N$, of binding elements such that

$$\mu_0 \xrightarrow{(t_1, b_1)} \mu_1, ..., \mu_{k-1} \xrightarrow{(t_k, b_k)} \mu_k.$$

We denote by $\mu_0 \xrightarrow{\sigma} \mu_k$ the fact that firing $\sigma$ yields the marking $\mu_k$. If

$$\mu_0 \xrightarrow{(t_1, b_1)} \mu_1 \xrightarrow{(t_2, b_2)} \mu_2 \xrightarrow{(t_3, b_3)} ...$$

for an infinite sequence $\sigma = (t_1, b_1) \cdot (t_2, b_2) \cdot (t_3, b_3)...$ then $\sigma$ is named an infinite occurrence sequence from $\mu_0$. A *reachable marking* of $BS$ is a marking which results from firing a finite occurrence sequence from $\mu_0$. The Boolean system $BS$ is *safe* iff every reachable marking of $BS$ marks each place of $BN$ with at most one token.

### 3.5 Definition *(Live, dead, synchronization deadlock)*

Consider a Boolean system $BS = (BN, \mu)$.

i) A binding element of $BS$ is *live* iff for every reachable marking $\mu_1$ of $BS$ the Boolean system $(BN, \mu_1)$ has a reachable marking which enables the given binding element. *BS* is *live with respect to all its high bindings* iff every high binding element of $BN$ is live.

ii) A transition of $t \in BS$ is *high-live* iff for every reachable marking $\mu_1$ of $BS$ the Boolean system $(BN, \mu_1)$ has a reachable marking, which enables at least one high binding element $(t,b), b \in B(t)$. $BS$ is *high-live* iff each transition is high-live.

iii) A transition $t \in BS$ is *dead* iff no reachable marking of $BS$ enables any binding element $(t,b), b \in B(t)$. The Boolean system $BS$ is *dead* iff all transitions of $BS$ are dead.

iv) Assume $BS$ to be safe. A marking $\mu_{dead}$ is named a *synchronization deadlock* for a Boolean transition $t \in BS$, if $skel(\mu_{dead})$ enables $t$, but no binding element $(t,b), b \in B(t)$, is enabled at $\mu_{dead}$. The Boolean system $BS$ is free from synchronization deadlocks iff no reachable marking is a synchronization deadlock.

Liveness of a binding element $(t,b), b \in B(t)$, is a much stronger condition than high-liveness of the corresponding transition $t \in BS$. Theorem 3.11 will make precise the equivalence of high-liveness and the absence of synchronization deadlocks. However, verifying that all binding elements of a transition are live requires more refined methods from model checking, see Proposition 4.8, ii). Our correctness criterion for Boolean systems is *well-behavedness* in the sense of Definition 3.6.

### 3.6 Definition *(Well-behavedness)*

A Boolean system $BS$ is *well-behaved* iff it is safe and live with respect to all its high bindings; otherwise $BS$ is named *ill-behaved*.

We will see that verification of safeness is the easy part. Any discrete Petri net morphism $PN_1 \xrightarrow{f} PN_2$ maps enabled occurrence sequences of $PN_1$ to enabled occurrence sequences of $PN_2$. The following Lemma 3.7 is a simple application of this fact.

### 3.7 Lemma *(Deriving saveness of a Boolean system)*

A Boolean system $BS$ is safe if its skeleton $BS^{skel}$ is safe.

**Proof.** Because the skeleton morphism

$$skel : BS \longrightarrow BS^{skel}$$

maps enabled occurrence sequences, it maps any reachable marking of $BS$ to a reachable marking of $BS^{skel}$. If no reachable marking of $BS^{skel}$ marks a place with more than a single token, the same holds true for $BS$, q. e. d.

The lifting problem considers the converse situation.

### 3.8 Definition *(Lifting property of a morphism)*

A Petri net morphism

$$PN_1 \xrightarrow{f} PN_2$$

has the *lifting property* iff for any enabled occurrence sequence $\sigma_2$ of $PN_2$ an enabled occurrence sequence $\sigma_1$ of $PN_1$ exists with $f(\sigma_1) = \sigma_2$. The occurrence sequence $\sigma_1$ is named a *lift of $\sigma_2$ against $f$*.

To decide if a morphism has the lifting property is not an easy task in general. For the skeleton of a Boolean system Lemma 3.9 solves the lifting problem.

### 3.9 Lemma *(Lifting property of the skeleton)*

If a safe Boolean system $BS = (BN, \mu)$ is free of synchronization deadlocks, then the skeleton morphism

$$skel : BS \longrightarrow BS^{skel}$$

has the lifting property and $BS^{skel}$ is safe too. In addition, the lift to high binding elements can be prescribed along an arbitrary path: Consider an enabled occurrence sequence $\sigma^{skel}$ from $BS^{skel}$ containing a sequence $t_0 \cdot ... \cdot t_{n-1}$ of transitions which extends to a path in $BS$

$$\gamma = (p_0, t_0, p_1, ..., t_{n-1}, p_n) \text{ with places } p_i, \ 0 \leq i \leq n,$$

and assume that the initial place $p_0$ is high-marked at $\mu$. Then $\sigma^{skel}$ has a lift $\sigma$ to $BS$ containing a sequence $(t_0, b_0) \cdot ... \cdot (t_{n-1}, b_{n-1})$ of high binding elements $(t_i, b_i)$, $0 \leq i < n$.

**Proof.** We may assume that $\sigma^{skel}$ is a single transition $t^{skel} \in BN^{skel}$ firing according to $skel(\mu) \xrightarrow{\sigma^{skel}} \mu_1^{skel}$. All pre-places of the corresponding transition $t \in BN$ are marked. Because $BS$ is free of synchronization deadlocks, the marking $\mu$ enables a binding $b \in B(t)$ of $BS$ with $\sigma^{skel} = skel(t, b)$. Due to the fairness of Boolean transitions, cf. Definition 3.3, the binding $b$ can be chosen according to the demand of $\gamma$. Therefore the occurrence sequence $\sigma := (t, b)$ of $BS$ is a suitable lift of $\sigma^{skel}$.

To prove safeness of $BS^{skel}$ we assume an occurrence sequence $\sigma^{skel}$ of $BS^{skel}$, which leads to a marking with more than one token at a given place. Lifting $\sigma^{skel}$ to an occurrence sequence $\sigma$ of $BS$ with the same property produces a contradiction to the safeness of $BS$, q. e. d.

A Boolean system $BS = (BN, \mu)$ is named *strongly connected* if the skeleton net $BN^{skel}$ is strongly connected. In this case $BN^{skel}$ is a T-system. This fact makes precise our former statement that Boolean systems generalize T-systems by adding the possibility of choice and representing the omission of actions by a second type of tokens.

### 3.10 Corollary (*Skeleton of a safe and high-live Boolean system*)

A safe and high-live Boolean system $BS$ is strongly connected. Its skeleton $BS^{skel}$ is live and safe.
**Proof**. High-liveness implies that $BS$ is free of synchronization deadlocks. According to Lemma 3.9 the skeleton is safe and the morphism $skel : BS \longrightarrow BS^{skel}$ has the lifting property: For any reachable marking $\mu^{skel}$ of $B^{skel}$ a reachable marking $\mu$ of $BS^{skel}$ exists with $skel(\mu) = \mu^{skel}$. High-liveness of $BS$ implies: For any transition $t \in B^{skel}$ a binding element $(t, b), b \in B(t)$, of $BN$ and an occurrence sequence $\mu \xrightarrow{\sigma} \mu_1$ exist with $\mu_1$ enabling $(t, b)$. As a consequence

$$skel(\mu) = \mu^{skel} \xrightarrow{skel(\sigma)} skel(\mu_1)$$

and $skel(\mu_1)$ enables $t$. As a consequence $BS^{skel}$ is live. According to Theor. 2.25 in [De1995] the underlying net of a live and safe Petri net is strongly connected, q. e. d.

Non-deadness of a bounded and strongly connected free-choice system implies its liveness, see

Theor. 4.31 in [DE1995]. We derive a similar property for Boolean systems as a consequence from the lifting property of the skeleton. Theorem 3.11 is a slight generalization of a result of Genrich and Thiagarajan who proved the statement for Boolean systems with only AND- or XOR-transitions, see Theor. 2.12 and Lemma 3.10 in [GT1984].

### 3.11 Theorem *(Liveness versus synchronization deadlock)*

For a Boolean system $BS$ with safe skeleton $BS^{skel}$ the following properties are equivalent:

1. $BS$ is high-live.
2. $BS$ is strongly connected and no reachable marking of $BS$ is dead.
3. $BS$ is free from synchronization deadlocks and the skeleton $BS^{skel}$ is live.

**Proof.** According to Lemma 3.7 the Boolean system $BS$ is safe. Set $BS = (BN, \mu_0)$.

$1 \Rightarrow 2$ Strong connectedness follows from Corollary 3.10, while high-liveness apparently implies non-deadness.

$2 \Rightarrow 3$ The assumption implies that at any reachable marking of $BS$ enables at least one binding element of $BN$. Therefore $BS$ has an infinite occurrence sequence $\sigma$ from $\mu_0$. It projects along $BS \xrightarrow{skel} BS^{skel}$ to an infinite occurrence sequence $\sigma^{skel}$ from $\mu_0^{skel}$. Therefore $BS^{skel}$ is live according to Theor. 3.17 in [De1995].

Assume the existence of a synchronization deadlock $\mu_{sd}$ of a transition $t \in BS$. By assumption $\mu_{sd}$ is not dead. Therefore an infinite occurrence sequence $\sigma$ from $\mu_{sd}$ exists. It projects to an infinite occurrence sequence $\sigma^{skel}$ from $skel(\mu_{sd})$, which avoids the distinguished transition $t \in BS^{skel}$. This fact is a contradiction, cf. the proof of Theor. 3.17 in [De1995].

$3 \Rightarrow 1$ Consider a reachable marking $\mu$ of $BS$ and a given transition $t$ of the underlying net. Because the initial marking of $BS$ contains at least one high-token, the same holds true for $\mu$. Therefore a transition $t_1$ exists with a pre-place high-marked at $\mu$. According to Theor. 1.14 and Theor. 1.15 in [GT1984] a minimal occurrence sequence $skel(\mu) \xrightarrow{\sigma_1^{skel}} \mu_1^{skel}$ of $BS^{skel}$ exists with $\mu_1^{skel}$ a blocking marking associated to $t_1$, i.e. $\mu_1^{skel}$ enables $t_1$ but no other transition of $BS^{skel}$. By Lemma 3.9 the occurrence sequence $\sigma_1^{skel}$ lifts to $\mu \xrightarrow{\sigma_1} \mu_1$ such that $\mu_1$ enables a high binding of $t_1$. Because $\mu_1^{skel}$ is a blocking marking, the live $T$-system $\left(BN^{skel}, \mu_1^{skel}\right)$ contains an unmarked path $\beta^{skel}$ from $skel(t_1)$ to $skel(t)$. A minimal occurrence sequence

$$\mu_1^{skel} \xrightarrow{\sigma_2^{skel}} \mu_2^{skel}$$

exists with $\mu_2^{skel}$ enabling $skel(t)$ and with the transitions from $\beta^{skel}$ contained as a subsequence in $\sigma_2^{skel}$. By Lemma 3.9 the occurrence sequence $\sigma_2^{skel}$ has a lift $\mu_1 \xrightarrow{\sigma_2} \mu_2$, such that $\mu_2$ enables a high binding of $t$, q. e. d.

## 4 Model checking of Boolean systems

In the present chapter we combine the theory of finite complete prefixes with the propositional logic of Boolean transitions to derive a model checking algorithm for Boolean systems $BS$, see Algorithm 4.9. Our investigation is based on the skeleton morphism

$$skel : BS \longrightarrow BS^{skel}$$

which has been introduced in Definition 3.1, iv). We first apply the prefix theory to $BS^{skel}$. This task is greatly facilitated by the fact that all places of $BS^{skel}$ are unbranched. As a consequence, all branching processes of $BS^{skel}$, in particular the unfolding and finite complete prefixes, are processes already.

### 4.1 Definition *(Colouring of a process and reachable base markings)*

Consider a Boolean system $BS = (BN, \mu_0)$ with skeleton

$$skel : BS \longrightarrow BS^{skel} = \left(BN^{skel}, \mu_0^{skel}\right)$$

and a finite process $pr : ON \longrightarrow BN^{skel}$ of the skeleton with

$$mark\,(\min(ON)) = mark\,(\max(ON)).$$

i) The *colouring of pr induced by BN* is a Boolean net $BON$ together with a morphism

$$cov : BON \longrightarrow BN.$$

The net $BON$ has the skeleton $ON = (B, E, K)$, Boolean transitions $skel^{-1}(pr(e))$, $e \in E$, and Boolean places $skel^{-1}(pr(b))$, $b \in B$. The morphism cov is induced by $pr$.

ii) Any marking $\mu$ of $BON$ with

$$supp(\mu) = \min(BON) \text{ and } skel(cov(\mu)) = skel(\mu_0)$$

is named a *base marking* of $BON$. A distinguished base marking $\mu_{0,\min}$ of $BON$ exists with $cov(\mu_{0,\min}) = \mu_0$.

iii) The set of *reachable base markings of BON originating from* $\mu_0$ is the smallest set *ReachBase* of base markings of $BON$ with the following properties:

- $\mu_{0,\min} \in ReachBase$
- If $\mu \in ReachBase$ and $\mu_1$ a base marking, which is reachable in $(BON, \mu)$, then also $\mu_1 \in ReachBase$.

For each reachable base marking $\mu \in ReachBase$ the induced morphism of Boolean systems

$$(BON, \mu) \longrightarrow (BN, cov(\mu))$$

is named the *base process starting from* $\mu$.

We consider the morphism $cov : BON \longrightarrow BN$ from Definition 4.1, i) a covering, but we will not provide a formal definition of this concept. The morphism fits into the commutative diagram from Figure 5. Readers used to category theory may take it as the definition of

$$cov : BON \longrightarrow BN,$$

that the diagram from Figure 5 is a fibre product in the category of Boolean nets:

$$\begin{array}{ccc} BON & \xrightarrow{cov} & BN \\ \downarrow skel & & \downarrow skel \\ ON & \xrightarrow{pr} & BN^{skel} \end{array}$$

*Figure 5: Finite complete prefix and its colouring*

On the basis of the morphisms from Figure 5 we explain our method of verification for a Boolean system $BS = (BN, \mu_0)$ with a live and safe skeleton.

- On the level of ordinary Petri nets: We choose for the skeleton a finite complete

prefix $ON \xrightarrow{pr} BN^{skel}$ with $skel(\mu_0) = mark(\min(ON)) = mark(\max(ON))$.

- On the level of Boolean systems: We determine $cov: BON \longrightarrow BN$ and all base processes $(BON, \mu) \longrightarrow (BN, cov(\mu))$, which start from an arbitrary reachable base marking $\mu$ of $BON$ originating from $\mu_0$.

- On the level of propositional logic: For each base process we employ model checking to find out the reachability of certain markings.

Because the Boolean transitions from $BON$ and $BN$ correspond bijectively to each other, we will not distinguish between the guard formula of a transition $t \in BON$ and the guard formula of the image transition $cov(t) \in BN$. As a consequence, we also do not distinguish between binding elements of $BON$ and $BN$.

For a finite occurrence net $ON$ we number its events according to the order from a topological sorting of the nodes of $ON$. The morphism $pr: ON \longrightarrow BN^{skel}$ transfers this numbering also to the transitions of $BN^{skel}$ and a posteriori to the transitions of the Boolean nets $BN$ and $BON$. In the following any numbering of transitions refers to this numbering.

## 4.2 **Example** *(Base process)*

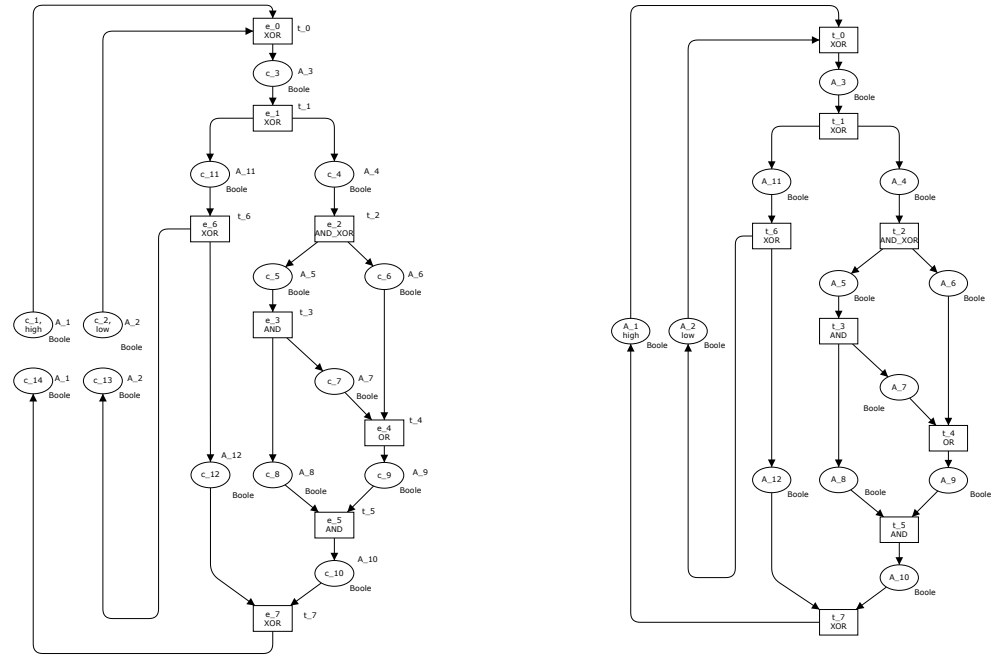

*Figure 6: A base process of the Boolean system from Figure 4*

Figure 6 (right hand side) repeats the Boolean system $BS = (BN, \mu)$ from Example 3.2, ii) and adds on the left hand side the Boolean system $(BON, \mu_{min})$. The figure illustrates from left to right the covering $cov: BON \longrightarrow BN$ and the induced base process

$$(BON, \mu_{min}) \longrightarrow (BN, \mu)$$

starting from $\mu_{min}$. The annotation of the nodes is similar to the annotation from Example 2.4.

Prefix theory of ordinary Petri nets has been recalled in Chapter 2. Our next step in the present chapter investigates the structure of all reachable base markings. Here fore we introduce the immediate successor relation and the successor graph of reachable base markings. When considering a Boolean system $BS = (BN, \mu_0)$ with a safe skeleton $BS^{skel}$ we will always assume in the following, that a covering $cov: BON \longrightarrow BN$ has been selected on the basis of a distinguished finite complete prefix $pr: ON \longrightarrow BN^{skel}$ of $BS^{skel}$ with

$$skel(\mu_0) = mark(\min(ON)) = mark(\max(ON)).$$

### 4.3     Definition *(The successor graph of reachable base markings)*

Consider a safe Boolean system $BS = (BN, \mu_0)$ and a distinguished covering $cov: BON \longrightarrow BN$. On the set of all reachable base markings of $BON$ we define the following partial order: $\mu_2 \in ReachBase$ is an *immediate successor* of $\mu_1 \in ReachBase$, iff $\mu_2$ is reachable in $(BON, \mu_1)$. The *successor graph* $\Gamma_{suc}$ of reachable base markings of $BON$ is the following directed graph: Vertices of $\Gamma_{suc}$ are the reachable base markings of $BON$ originating from $\mu_0$, an oriented arc from $\mu_1$ to $\mu_2$ exists in $\Gamma_{suc}$ iff $\mu_2$ is an immediate successor of $\mu_1$.

The graph $\Gamma_{suc}$ is connected. The reason is: Each reachable base marking $\mu$ can be reached from $\mu_{0,\min}$ along a finite path - by construction. In general $\Gamma_{suc}$ is not strongly connected. We denote by $\Gamma C_{suc}$ the induced graph of strong components of $\Gamma_{suc}$. For each vertex $x \in \Gamma_{suc}$ the class $[x] \in \Gamma C_{suc}$ denotes the strong component of $x$ as a set of vertices from $\Gamma_{suc}$. By construction the directed graph $\Gamma C_{suc}$ is acyclic. Its oriented arcs define a partial order on the vertices of $\Gamma C_{suc}$. The order has the unique minimal element $\min \Gamma C_{suc} = \{[\mu_0]\}$ and a finite set

$$\max \Gamma C_{suc} = \{cp_1, ..., cp_n\}$$

of maximal elements. Elements of $\min \Gamma C_{suc}$ and $\max \Gamma C_{suc}$ are named respectively *minimal* and *maximal* strong components of $\Gamma_{suc}$.

For the Boolean system $BS = (BN, \mu)$ from our running Example 3.2 the coloured net $BON$ from Example 4.2 has two reachable base markings

$$\mu_1 := \mu \text{ with } Mark(\mu_1) = A_1 \wedge (\neg A_2) \text{ and } \mu_2 \text{ with } Mark(\mu_2) = (\neg A_1) \wedge A_2.$$

We have $\mu_1 \leq \mu_2$ and $\mu_2 \leq \mu_1$. Therefore, the successor graph $\Gamma_{suc}$ has only two vertices and is strongly connected. Its unique strong component $cp = \{\mu_1, \mu_2\}$ is both minimal and maximal.

The concept of the successor graph of reachable base markings permits us to split arbitrary occurrence sequences of a safe Boolean system $BS$ into occurrence sequences of fixed length. Their length depends only on the choice of a finite complete prefix of an unfolding of $BS^{skel}$. Each of these fragmented occurrence sequences of bounded length can be studied with one of the base processes.

### 4.4     Remark *(Permutation of occurrence sequences)*

We recall that an enabled transition of a T-system loses its firing concession only by firing itself. This fact applies to the skeleton of a Boolean system $BS$ and generalizes to binding elements of $BS$: Consider a fixed Boolean transition $t \in BS$. An enabled binding element $(t,b) \in B(t)$ loses its concession only by firing itself or by the firing of another enabled binding element of the same transition $t$. As a consequence, Lemma 3.24 in [DE1995] about the permutation of occurrence sequences in T-systems applies mutatis mutandis also to occurrence sequences of $BS$.

### 4.5 Lemma *(Characterization of reachability)*

Consider a safe Boolean system $BS = (BN, \mu_0)$ and a distinguished covering $\text{cov}: BON \longrightarrow BN$. For any marking $\mu$ of $BN$ we have the equivalence:

i) The marking $\mu$ is reachable in $BS$.

ii) A reachable base marking $\mu_b$ of $BON$ originating from $\mu_0$ and a reachable marking $\tilde{\mu}$ of $(BON, \mu_b)$ with $\text{cov}(\tilde{\mu}) = \mu$ exist.

**Proof.** Because of $mark(\min ON) = mark(\max ON)$ the sum of all transitions $t \in BN^{skel}$ which occur due to the events from $ON$ is a T-invariant $\tau$. Here the transitions are taken with the multiplicity of their occurrence. All transitions from $BN^{skel}$ are contained in $\tau$ with the same multiplicity $k \in N^*$, i.e.

$$\tau = k \cdot \sum_{t \in BN^{skel}} t .$$

ii) => i) Because $\mu_b$ is a reachable base marking, a sequence $\mu_{b,i}, i = 0,...,n$, of reachable base markings exists with

$\mu_{b,0} = \mu_{0,\min}$, $\mu_{b,i+1}$ is an immediate successor of $\mu_{b,i}$ for $i = 0,...,n-1$, $\mu_{b,n} = \mu_b$.

The corresponding base processes $(BON, \mu_{b,i}) \longrightarrow (BN, \text{cov}(\mu_{b,i}))$ induce in $BN$ occurrence sequences

$$\mu_0 \longrightarrow \text{cov}(\mu_{b,1}) \longrightarrow ... \longrightarrow \text{cov}(\mu_{b,n}) .$$

And the assumed reachability of $\tilde{\mu}$ in $(BON, \mu_b)$ provides an occurrence sequence

$$\text{cov}(\mu_b) \longrightarrow \text{cov}(\tilde{\mu}) = \mu .$$

The catenation of all occurrence sequences from above is an occurrence sequence of $BS$ which leads to $\mu$.

i) => ii) Because $ON$ is a complete prefix, a cut $c$ of $ON$ exists with $mark(c) = skel(\mu)$. Due to Remark 4.4 we may permute occurrence sequences from $BS$ which lead to $\mu$. Multiple application of Lemma 3.24 in [DE1995] and consideration of the special form of $\tau$ provide

- a series $\mu_i \xrightarrow{\sigma_i} \mu_{i+1}, i = 0,...,n-1$, of occurrence sequences of $BN$ such that each $\sigma_i$, $i = 0,...,n-1$, fires exactly $k$ binding elements of each transition $t \in BN$

- and an occurrence sequence $\mu_n \xrightarrow{\sigma_n} \mu$ firing no more than $k$ binding elements of each transition $t \in BN$ and less than $k$ binding elements for at least one transition from $BN$.

As a consequence we obtain reachable base markings $\mu_{i,\min}, i = 0,...,n$, with each $\mu_{i+1,\min}$ an immediate successor of $\mu_{i,\min}$. With $\mu_b := \mu_{n,\min}$ we obtain $\tilde{\mu}$ a as reachable marking of $(BON, \mu_b)$, q. e. d.

Preparing our model checking algorithm we now attach to $\text{cov}: BON \longrightarrow BN$ a series of formulas from propositional logic, which represent certain behavioural properties of the system, see Table 2 and Definition 4.6. The satisfiability of these formulas is equivalent to the presence or absence of these properties. The column "Property" in Table 2 looks ahead to Proposition 4.8.

| Name | Definition | Context | Property |
|---|---|---|---|
| Marking formula | $Mark(\mu) :=$ $X_1 \wedge ... \wedge X_r \wedge \neg Y_1 \wedge ... \wedge \neg Y_s$ | arbitrary $BN$ | - |
| Reachability formula | $Reach(\mu_1) :=$ $Mark(\mu) \wedge G_0 \wedge ... \wedge G_i$ | $(BON, \mu)$ | $\mu_1$ reachable iff $Reach(\mu_1)$ satisfiable |
| Enabling formula | $Enabl(t,b) := Reach(\mu_{enabl}(t,b))$ | $(BON, \mu)$ | $(t,b)$ non dead iff $Enabl(t,b)$ satisfiable |
| Deadlock formula | $Dead(t) :=$ $Reach(\mu_{dead,1}) \vee ... \vee Reach(\mu_{dead,n})$ | $(BON, \mu)$ with safe $BS$ | synchronization deadlock of $t$ reachable iff $Dead(t)$ satisfiable |

*Table 2: Formulas representing the semantics of a Boolean system*

### 4.6 Definition *(Formulas representing the semantics of a Boolean system)*

i) Consider a Boolean net $BN$ with arc annotation $X : F \longrightarrow Var(BOOLE)$ and Boolean transitions $(t, G)$. For a marking $\mu$, with marks each place of $BN$ with at most one token, we define its *marking formula* as

$$Mark(\mu) := X_1 \wedge ... \wedge X_r \wedge \neg Y_1 \wedge ... \wedge \neg Y_s$$

with $\{X_1, ..., X_r\} = \{X(p,t) : p \in BN, t \in BN, \mu(p) = true\}$

and $\{Y_1, ..., Y_s\} = \{X(p,t) : p \in BN, t \in BN, \mu(p) = false\}$

the set of variables annotating the arcs starting or ending at marked places.

ii) Consider a safe Boolean system $BS = (BN, \mu_0)$, a distinguished covering $cov : BON \longrightarrow BN$ and a reachable base marking $\mu$ of $BON$ originating from $\mu_0$.

- For a marking $\mu_1$ of $BON$ we define its *reachability formula* with respect to $(BON, \mu)$ as

$$Reach(\mu_1) := Mark(\mu) \wedge G_0 \wedge ... \wedge G_i$$

  with $i$ the minimal index such that firing the occurrence sequence $\sigma = t_0 ... t_i$ of the skeleton $BS^{skel}$ creates the marking $skel(cov(\mu_1))$.

- For a binding element $(t,b), b \in B(t)$, of a Boolean transition $t$ of $BN$ we define its *enabling formula* with respect to $(BON, \mu)$ as

$$Enabl(t,b) := Reach(\mu_{enabl}(t,b))$$

  with $\mu_{enabl}(t,b)$ the enabling marking of $(t,b)$ considered as a binding element of $BON$.

- For a Boolean transition $t$ of $BN$ we define its *deadlock formula* with respect to $(BON, \mu)$ as

$$Dead(t) := Reach(\mu_{dead,1}) \vee ... \vee Reach(\mu_{dead,n})$$

  with $\mu_{dead,1}, ..., \mu_{dead,n}$ the synchronization deadlocks of $t$ considered as a transition of $BON$.

### 4.7 Remark *(Reachability and satisfiability)*

With the notations of Definition 4.6: According to Lemma 4.5 the reachability of $\text{cov}(\mu_1)$ in the original Boolean system $BS$ is equivalent to the reachability of $\mu_1$ in $(BON, \mu_{pre})$. And $\mu_1$ is reachable in $(BON, \mu_{pre})$ iff its reachability formula $Reach(\mu_1)$ with respect to $(BON, \mu_{pre})$ is satisfiable.

Apparently, binary opening transitions do not have any synchronization deadlocks. As a consequence their deadlock formula is the constant *false*. Table 3 shows the deadlock formulas of binary closing Boolean transitions of different logical type. Their arc annotations refer to Figure 3. The column "Distinguished enabling formula(s)" will be referred to when explaining Algorithm 4.9.

| Logical type | Deadlock formula | Distinguished enabling formula(s) |
|---|---|---|
| AND | $Reach(X \wedge \neg Y) \vee Reach(\neg X \wedge Y)$ | - |
| XOR | $Reach(X \wedge Y)$ | - |
| AND_XOR | $Reach(\neg X \wedge Y)$ | $Reach(X \wedge \neg Y)$ |
| XOR_AND | $Reach(X \wedge \neg Y)$ | $Reach(\neg X \wedge Y)$ |
| OR | *false* | $Reach(X \wedge Y)$, $Reach(X \wedge \neg Y)$, $Reach(\neg X \wedge Y)$ |

*Table 3: Deadlock formulas and distinguished enabling formulas of binary closing Boolean transitions*

### 4.8 Proposition *(Model checking of a safe Boolean system)*

Consider a safe Boolean system $BS = (BN, \mu_0)$ and a distinguished covering

$$\text{cov}: BON \longrightarrow BN .$$

i) $BS$ is free from synchronization deadlocks iff for all Boolean transitions $t \in BN$ and each reachable base marking $\mu$ of $BON$ the formula $Dead(t)$ with respect to $(BON, \mu)$ is not satisfiable.

ii) If $BS$ is free from synchronization deadlocks, then $BS$ is live with respect to all its high-bindings iff for each transition $t \in BS$ and each high binding element $(t,b), b \in B(t)$ the following holds true: Each maximal strong component $cp \in \Gamma C_{suc}$ of the successor graph $\Gamma_{suc}$ of $BS$ contains a reachable base marking $\mu \in cp$, such that the formula $Enabl(t,b)$ with respect to $(BON, \mu)$ is satisfiable.

**Proof**. ad i) Because $Dead(t)$ is a disjunction of reachability formulas, it is satisfiable iff at least one of these reachability formulas is satisfiable. A reachability formula like

$$Reach(\mu_{dead}) := Mark(\mu) \wedge G_0 \wedge \ldots \wedge G_i$$

with respect to $(BON, \mu)$ is satisfiable iff $\mu_{dead}$ is reachable in $(BON, \mu)$. Therefore the statement follows from Remark 4.7.

ad ii) We assume that the binding $(t,b)$ is live. We consider an arbitrary strong component $cp \in \max \Gamma C_{suc}$ and an arbitrary reachable base marking $\mu_b \in cp$. By assumption a reachable marking $\mu_1$ of $BS_b := (BN, \operatorname{cov}(\mu_b))$ exists which enables $(t,b)$. Applying Lemma 4.5 to the Boolean system $BS_b$ we obtain a reachable base marking $\mu$ of $BON$ originating from $\operatorname{cov}(\mu_b)$ and a reachable marking $\tilde{\mu}_1$ of $(BON, \mu)$ with $\operatorname{cov}(\tilde{\mu}_1) = \mu_1$. Note that $\mu$ is also a reachable base marking originating from $\mu_0$. Due to the maximality of the strong component $cp$ each reachable base marking originating from $\operatorname{cov}(\mu_b)$ belongs to $cp$. Therefore $\mu \in cp$. The statement now follows from Remark 4.7.

For the opposite direction we have to prove that the binding element $(t,b)$ is live. We start with a reachable marking $\mu_{pre}$ of $BS$ and have to find a reachable marking $\mu_{post}$ of $(BN, \mu_{pre})$ which enables $(t,b)$. According to Lemma 4.5 a reachable base marking $\mu$ of $BON$ and a reachable marking $\tilde{\mu}_{pre}$ of $(BON, \mu)$ with $\operatorname{cov}(\tilde{\mu}_{pre}) = \mu_{pre}$ exist. Like $BS$ also $(BON, \mu)$ is free from synchronization deadlocks. According to Lemma 3.9 at least one immediate successor $\mu_1$ of $\mu$ is reachable in $(BON, \tilde{\mu}_{pre})$. In $\Gamma C_{suc}$ a path exists from $[\mu_1]$ to a maximal element $cp \in \max \Gamma C_{suc}$. By assumption a base marking $\mu_2 \in cp$ and a reachable marking $\mu_3$ of $(BON, \mu_2)$ exist such that $\mu_2$ enables $(t,b)$ when considered a binding element of $BON$. The marking $\mu_{post} := \operatorname{cov}(\mu_2)$ is reachable in $(BN, \mu_{pre})$ by construction and enables $(t,b)$ when the latter is considered a binding element of $BN$, q. e. d.

The main result of the paper is the following Algorithm 4.9 for the verification of a Boolean system with a live and safe skeleton.

### 4.9    Algorithm *(Liveness of a safe Boolean system)*

Input: Binary Boolean system $BS = (BN, \mu_0)$ with a live and safe skeleton $BS^{skel}$.

Output:

- List of Boolean transitions of $BS$ which suffer a synchronizing deadlock.
- List of transitions of $BS$ which are not live with respect to all their high bindings.

| | |
|---|---|
| | Determine a finite complete prefix $ON \xrightarrow{pr} BN^{skel}$ of the unfolding of $BS^{skel}$ with $pr(\max ON) = pr(\min ON)$ |
| | Set $\operatorname{cov}: BON \longrightarrow BN$ the colouring of $pr: ON \longrightarrow BN^{skel}$ |
| | Determine the successor graph $\Gamma_{suc}$ of reachable base markings originating from $\mu_0$ |
| | Is $(BON, \mu)$ deadlock free for each reachable base marking $\mu \in \Gamma_{suc}$? |
| No | Yes |
| - | Determine the set $\max \Gamma C_{suc}$ of maximal strong components of $\Gamma_{suc}$ |
| | For each $cp \in \max \Gamma C_{suc}$ and each high binding $(t,b)$ of $BN$ check, whether for at least one $\mu \in cp$ the binding element $(t,b)$ can be enabled in $(BON, \mu)$ |
| | Output result |

We will now discuss in detail the different steps of algorithm 4.9.

i) The T-system $BS^{skel} = (BN^{skel}, \mu_0^{skel})$ is live and safe iff every circuit of $BN^{skel}$ is marked at $\mu_0^{skel}$ and every place of $BN^{skel}$ is contained in a circuit marked at $\mu_0^{skel}$ with exactly one token, see Theor. 3.15 and 3.18 in [DE1995].

ii) For all examples in the present paper the algorithm from [ERV2002] constructs a finite complete prefix $ON \longrightarrow BN^{skel}$ with

$$pr(\max ON) = pr(\min ON).$$

However, this equality does not hold for any minimal finite complete prefix.

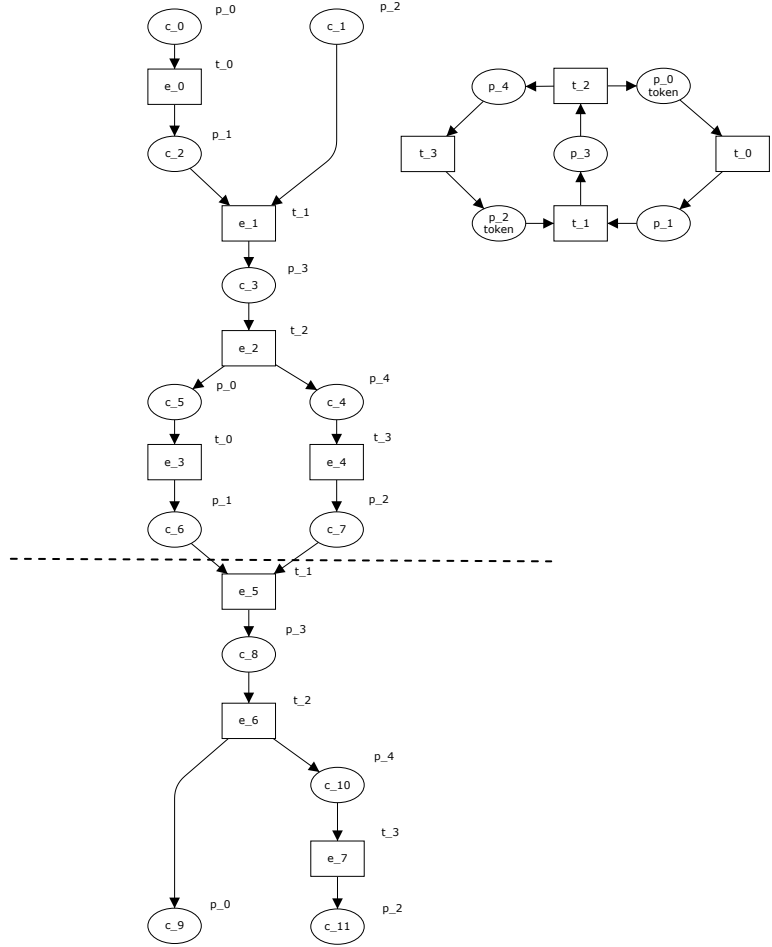

*Figure 7: Finite complete prefixes of a live and safe T-system (right)*

Figure 7 shows a finite complete prefix $pr_2 : ON_2 \longrightarrow N$ of the unfolding of a live and safe T-system $(N, \mu_0)$. The part of Figure 7 above the dotted line shows the restriction

$$pr_1 : ON_1 \longrightarrow N$$

with $ON_1 \subset ON_2$ the subnet generated by the nodes $\{c_0,...,c_7, e_0,...,e_4\}$. It is a finite complete prefix too. Its final cut $\max ON_1 = \{c_6, c_7\} \subset ON_1$ satisfies

$$pr(\max ON_1) = \{p_1, p_2\} \neq pr(\min ON_1) = \{p_0, p_2\}$$

and does not restore the original marking, i.e., $mark(\max ON_1) \neq \mu_0$.

Because a live and bounded T-system is cyclic, each finite process $pr_1$ of $BS^{skel} = \left(BN^{skel}, \mu_0^{skel}\right)$ extends to a finite process $pr_2$, $pr_1 \subseteq pr_2$, which restores the original marking $\mu_0^{skel}$. We may assume that $pr_2$ can be obtained from $pr_1$ without firing all transitions of $BN^{skel}$, see Lemma 3.24 in [DE1995]. As a consequence: If $t_0 \in BN^{skel}$ is a transition with maximal occurrence

$$n := \left| pr_1^{-1}(t_0) \right| = \max\left\{ \left| pr_1^{-1}(t) \right| : t \in BN^{skel} \right\}$$

then

$$n = \left| pr_2^{-1}(t) \right| \text{ for all } t \in BN^{skel}.$$

In the example from Figure 7 we have $n = 2$.

iii) To obtain the colouring $cov : BON \longrightarrow BN$ of the process $pr : ON \longrightarrow BN^{skel}$ one has to annotate each place of the occurrence net $ON$ with the set $Boole$, each arc with the same variable as the corresponding arc in $BN$ and each transition of $ON$ with the same guard function as the corresponding transition in $BN$, see Definition 4.1, i).

iv) For the computation of the successor graph $\Gamma_{suc}$ of reachable base markings Algorithm 4.9 first computes the set $MB$ of base markings of $BON$. The markings from $MB$ correspond to the different combinations of high and low tokens on the places of $min(BON)$.

In the next step the algorithm determines which of the base markings from $MB$ are reachable base markings. Most simple is the case of a singleton $min(BON) = \{ c \}$. It corresponds to the case that only one place $p = cov(c)$ of $BN$ is marked at the initial marking $\mu_0$: By Definition 3.1 the token content of $p$ is a high-token and due to the fairness condition in Definition 3.3 the other marking from $MB$ is not reachable because it does not comprise any high token. Therefore the marking $\mu_{0,\min}$ is the only reachable base marking.

In general Algorithm 4.9 employs model checking to determine which of the markings $\mu \in MB$ are reachable. The construction of $\Gamma_{suc}$ may proceed with the iterative construction of a spanning tree $\Gamma T$. Starting with $\Gamma T = \{ \mu_{0,\min} \}$ one checks for each marking $\mu \in MB - \{ \mu_0 \}$ the reachability of $\mu$ in $(BON, \mu_{0,\min})$. If $\mu$ is reachable, then it is added to $\Gamma T$ and removed from $MB$. Iteratively the subsequent steps check the reachability of the remaining elements from $MB$ with respect to the already existing vertices from $\Gamma T$. As noted subsequent to Definition 4.6, a marking $\mu$ is reachable in $(BON, \mu_1)$ iff the formula $Reach(\mu)$ with respect to $(BON, \mu_1)$ is satisfiable.

v) Only for closing Boolean transitions of logical type AND, XOR, AND_XOR and XOR_AND (see Table 3) Algorithm 4.9 has to investigate possible synchronization deadlocks. The check is performed as satisfiability check according to Proposition 4.8, i).

vi) After verifying that the Boolean system $BS$ is free of synchronization deadlocks we can apply the lifting Lemma 3.9. As a consequence, from each reachable marking of $BS$ a marking is reachable, which marks the pre-place of a given opening transition of $BS$ with a high token. The marking therefore enables all high-bindings elements of the transition.

Similarly, each high binding element of a closing transition of logical type AND or XOR is live: Always a marking is reachable, which marks a given pre-place of the transition with a high-token, the other pre-place with a second token and such that the transition is not in a synchronization deadlock.

Only for closing transitions of logical type AND_XOR, XOR_AND and OR a separate investigation is needed. Table 3 shows those enabling formulas $Enabl(t,b)$ which Algorithm 4.9 has to check according to Proposition 4.8, ii).

# 5 Application to EPCs

One of the first questions, which comes up when checking a given EPC for correctness, is:

- Which are the boundary events of the EPC?

An EPC must be either without any boundary nodes or it must be bounded by events, having at least one in-event and one out-event. That's a syntactic property which can be easily checked. In-events are initial or triggering events. But in case of loops also inner initial events may exist. Similarly are out-events the terminal or goal events of the process. And in case of loops also inner terminal events may exist. The situation becomes more difficult when the EPC has more than one single in-event or more than one single out-event. In that case the second question is:

- Which combinations of in-events and which combinations of out-events are intended by the modeller of the EPC?

This question can no longer be answered by a syntactical analysis. In general it cannot even be answered by a semantical analysis. Instead the answer must be known before any semantical analysis can start. Sometimes the boundary events of the EPC are annotated by process indicators referring to processes at the next higher level of a hierarchical process model. Then the possible combinations of the boundary events derive top-down from the possible event combinations within the process model one level higher. But often such a model is lacking. To clarify the intention of the modeller in this case, one can use an algorithm to generate a proposal for the possible event combinations. The algorithm applies mirror reflexion to both the first logical connectors after the in-events and the last logical connectors before the out-events. However, if the modeller is not a hand and his intention cannot be read off from the name of the events, the model checker himself has to make an educated guess.

After these two questions have been answered, the verification of the EPC continues with adding a start/end-connection: We introduce a separate event „start/end" and connect this distinguished event by arcs and logical connectors to all intended combinations of start events. Similar we connect all intended combinations of end-events by logical connectors and arcs with the distinguished event. After this kind of short-circuiting the resulting EPC should be strongly connected. Otherwise the structure of the EPC is considered to be faulty. All following steps of the verification will presuppose a short-circuited, strongly connected EPC.

Most easy is the verification of AND/XOR EPCs. These are EPCs with logical connectors of type AND or XOR only. To define the semantics of AND/XOR-EPCs and for their verification no Boolean systems are necessary. Instead the EPC translates to a free-choice system: Events translate to places, functions to transitions, while logical connectors of type AND translate to transitions and logical connector of type XOR translate to places. Possibly some additional unbranched places or transitions have to be introduced for syntactical reasons. Each place which represents a start event gets marked with a token. The resulting ordinary Petri net is a free-choice system. It defines the free-choice semantics of the EPC [Aal1999]. Algorithms to verify liveness and safeness of free-choice systems resulting from AND/XOR EPCs are well-established, see Theor. 4.2 in [ES1990] or Theor. 5.8 in [DE1995].

Of course the free-choice semantics of an AND/XOR-EPC can also be obtained from its Boolean semantics, which results from translating the EPC into a Boolean system: Events translate to Boolean places, functions into unary Boolean transitions, while logical connectors of type AND and XOR translate to Boolean transitions of the corresponding logical type. Each in-event produces a high token on the corresponding place. In addition low tokens have to be added such that the skeleton is live and safe. If the resulting Boolean system $BS$ is restricted to the flow of high tokens then the free-choice system $BS^{high}$, which defines the free-choice semantics, is obtained, see [SW2010].

Now we address general EPCs with logical connectors of arbitrary type. The type may differ from AND or XOR.

## 5.1 Example *(Closing OR-connector)*

The example from Figure 8 shows an EPC with a closing OR-connector.

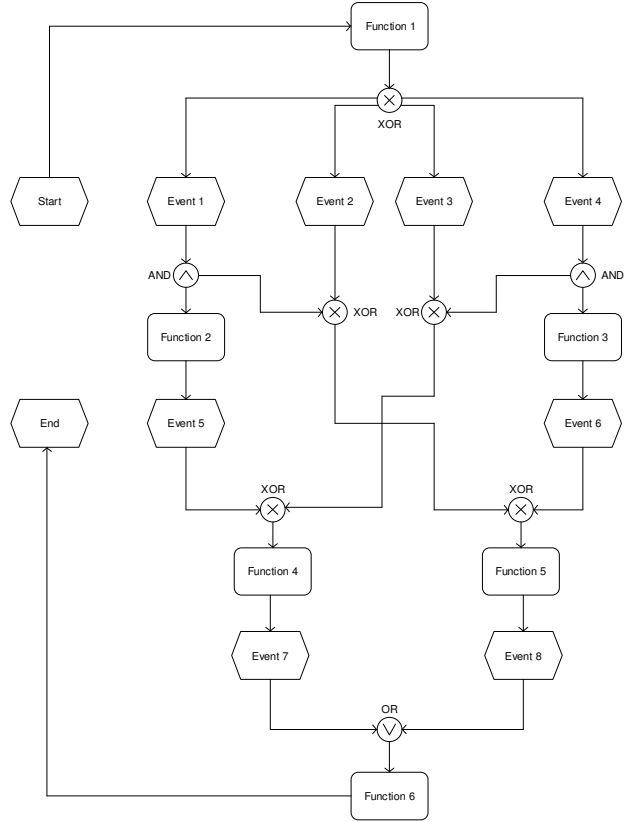

*Figure 8: EPC with closing OR-connector*

The process is triggered by the event *Start*. Depending on the outcome of *Function 1* exactly one of the events *Event 1*,…, *Event 4* happens. *Event 1* triggers both, *Function 2* and *Function 5*, while *Event 2* triggers only *Function 5*. Analogously *Event 4* triggers *Function 3* and *Function 4*, while *Event 3* triggers only *Function 4*. Finally the outcome of any combination of *Function 4* and *Function 5* triggers *Function 6* and the final event *End*. The process comprises a set of functions and events, which are activated according to a non-trivial control flow. Its logic is determined by a series of connectors of logical type AND, XOR and OR.

The modeller has provided the EPC with a single in-event and a single out-event. Therefore it is straightforward for the model checker to short-circuit the EPC.

Due to the OR-join the EPC from Figure 8 does not translate to a free-choice system as long as the difference between XOR and OR is respected. However, after translating the OR-connector to a Boolean transition of logical type OR the Boolean semantics of this EPC is well-defined. In addition Algorithm 4.9 verifies that the resulting Boolean system is well-behaved.

We are now returning to our running example „Loan request" from Figure 1. We collect all steps for its verification that we have developed in this paper.

### 5.2   **Example** *(Loan request)*

The EPC „Loan request" from Figure 1 has a logical connector of type different from AND or XOR. The verification of the EPC proceeds along the following steps:

- Translation of the EPC into a strongly-connected Boolean system $BS = (BN, \mu)$, see Figure 4.

- Verifying that the skeleton $BS^{skel}$ is safe and live.

- Applying Algorithm 4.9:

    o   Determination of a finite complete prefix $pr : ON \longrightarrow BN^{skel}$, see Figure 2.

- o Determination of the colouring $cov: BON \longrightarrow BN$ of $pr$, compare Figure 6.

- o Determination of the successor graph $\Gamma_{suc}$, see remark after Definition 4.3.

- o Applying Proposition 4.8 to the base processes $(BON, \mu_b) \longrightarrow (BN, cov(\mu_b))$.

Algorithm 4.9 outputs that the Boolean system from Figure 4 is ill-behaved:

```
Example 7: Loan request with loop
OK, no deadlock
Successor graph: 1 strong component(s):
Minimal - Maximal - Component 1 with 2 node(s):
 nicht A2 und A1  , A2 und nicht A1  ,
ERROR, non high-live transition(s):
Boolean transition of type: CLOSING OR with arc-variables: A6, A7, A9
```

Table 4: Model checking: EPC NOK

The Boolean system is deadlock free, but its closing Boolean transition $t_4$ of logical type OR is not live with respect to all its high-bindings. After changing its logical type to AND_XOR the Boolean system becomes well-behaved, see Table 5.

```
Example 21: Loan request with loop, corrected
OK, no deadlock
Successor graph: 1 strong component(s):
Minimal - Maximal - Component 1 with 2 node(s):
 nicht A2 und A1  , A2 und nicht A1  ,
OK, high-live
```

Table 5: Model checking: EPC OK

**Note**. In Table 4 and Table 5 the meaning of the German words is "nicht = not", "und = and".

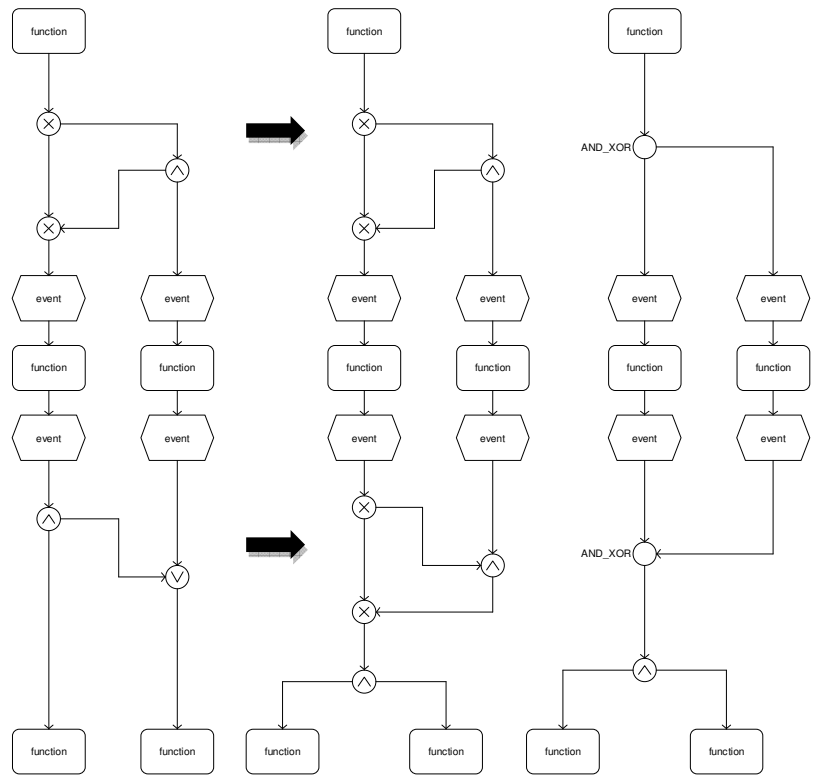

Figure 9: Different modelling constructs for the EPC "Loan request"

The authors of [MA2008] have employed a subtle logical construct to provide the EPC with an opening AND_XOR-connector. They used a combination of two XOR- and one opening AND-connector, Figure 9 (left hand side). To close the alternative the authors did not use the formally analogous construction with two XOR- and one closing AND-connector, Figure 9 (middle). The closing construction would have been erroneous, because it does not synchronize the decisions made at the indicated two XOR-splits in Figure 9 (middle). Instead the authors use one OR-join to close the AND_XOR alternative Figure 9 (left hand side). But one of the three firing modes of the closing OR will never be activated.

Different from the authors of [MA2008] we therefore consider the EPC from Figure 4 ill-behaved. In accordance with the above model checking result we propose to model the EPC with a pair of AND_XOR-alternatives like in Figure 9 (right hand side).

As a final example we consider an EPC from [DVV2006] proposed by the authors as a visualization of their method of EPC verification.

### 5.3   Example *(EPC describing the EPC verification process)*

The EPC from Figure 10 illustrates the verification procedure from [DVV2006] and reproduces Fig. 2 from [DVV2006]. The EPC exemplifies the difficulty for the model checker to short-circuit a given EPC. What are the intended initial events, which are the intended final events of the EPC from Figure 10?

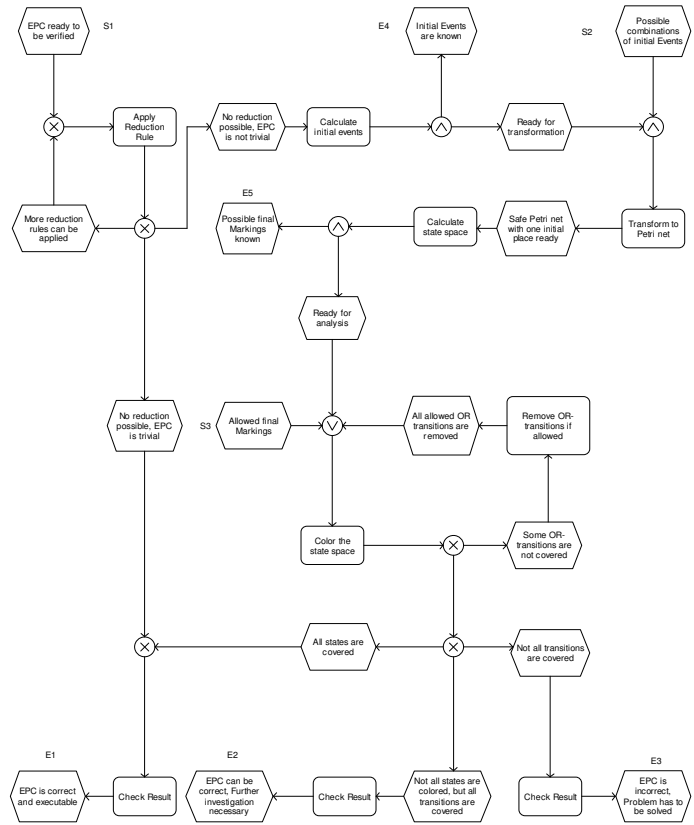

*Figure 10: EPC describing the EPC verification process, see [DVV2006]*

The EPC has three in-events "EPC ready to be verified" (S1), "Possible combinations of initial Events" (S2) and "Allowed final Markings" (S3) as well as five out-events "EPC is correct and executable" (E1), "EPC can be correct. Further investigation necessary" (E2), "EPC is incorrect, Problem has to be solved" (E3), "Initial Events are known" (E4) and "Possible final Markings known" (E5). From their annotation the reader cannot read off all intended combinations. But the translation of the EPC to a Petri net in [DVV2006], Fig. 3, achieved by the modellers themselves shows, that surprisingly some of these events are not intended as boundary events at all. The pair of events E4 and S2 as well as E5 and S3 seem to be annotations intended as hints for the human reader. The two components of each pair should therefore better be linked by two functions "t" and "u": Therefore we assume that the EPC intended by the modeller looks like Figure 11. It has a single in-event (S1) and an XOR-combination of the three out-events E1, E2 and E3.

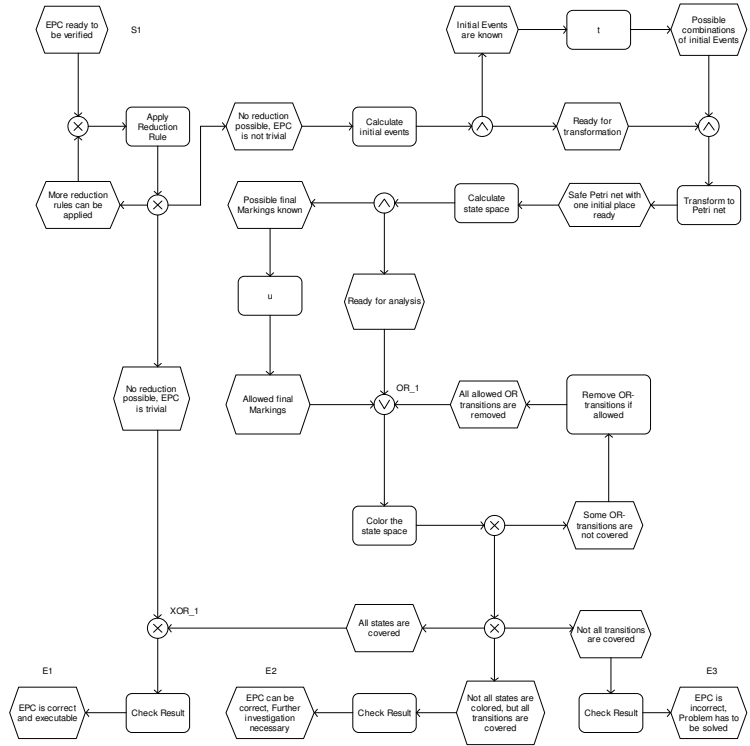

*Figure 11: Intended EPC describing the EPC verification process*

The OR-connector named $OR_1$ in Figure 11 decomposes into two binary OR-connectors as shown in Figure 12.

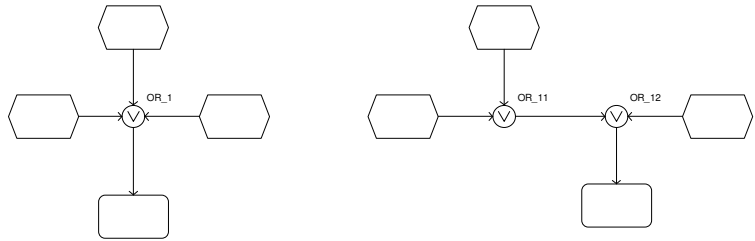

*Figure 12: Decomposition of the OR-connector $OR_1$ from Figure 11*

When the authors from [DVV2006] translate the EPC to a resembling Petri net, they transform the OR-join $OR_{11}$ in Figure 12 to a closing transition and the OR-join $OR_{12}$ to a closing place. And they transform the XOR-join $XOR_1$ from Figure 11 to a closing place too. The authors do not give any justification for these transformations. In particular they do not explain why they skip the difference between the OR-join and the XOR-join.

Figure 13 shows the binary Boolean system which results from short-circuiting and translating the EPC from Figure 11. Some unary transitions have been skipped in order to focus on the control flow.

The final events E1, E2 and E3 of Figure 11 translate to the places A_19, A_20 and A_21 of Figure 13. The initial event S1 translates to the marked place at the beginning of arc A_0. The two OR-joins $OR_{11}$ and $OR_{12}$ from Figure 12 translate respectively to the Boolean transitions $t_6$ and $t_7$ of logical type OR, while the XOR-join $XOR_1$ from Figure 11 translates to the Boolean transition $t_{10}$ of logical type XOR.

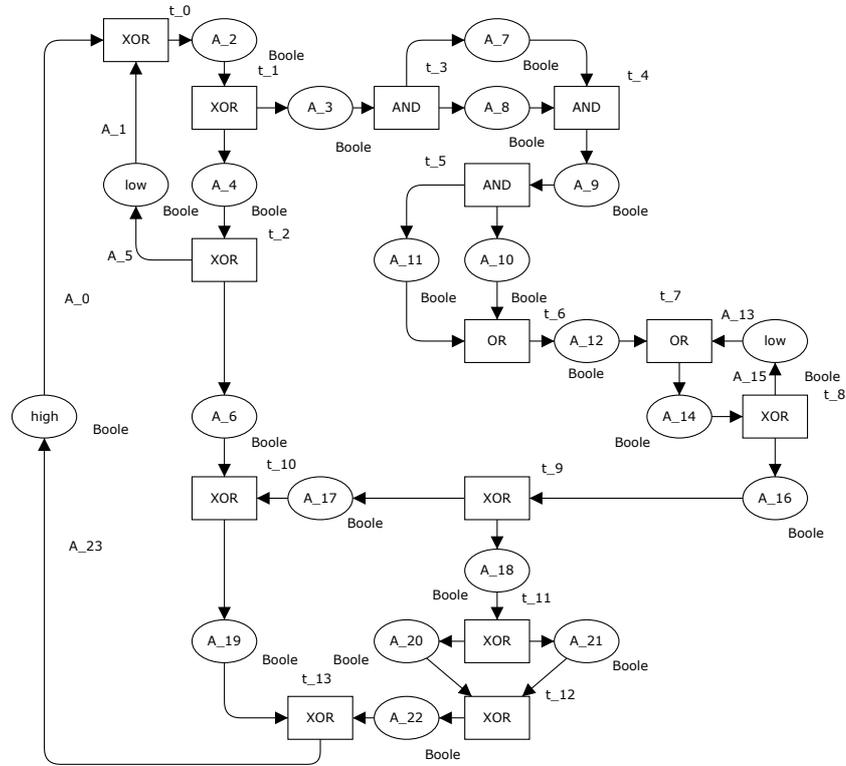

*Figure 13: Binary Boolean system of the EPC from Figure 11*

Algorithm 4.9 outputs that the Boolean system from Figure 13 is ill-behaved, see Table 6: The Boolean system is deadlock-free but the closing Boolean transitions $t_6$ and $t_7$ of logical type OR are not high-live with respect to all their bindings.

```
Example 38: van der Aalst et al.: Verification of the SAP Reference Models ... 2006
OK, no deadlock
Successor graph: 1 strong component(s):
Minimal - Maximal - Component 1 with 3 node(s):
 nicht A1 und  nicht A13 und A0  ,  nicht A1 und A13 und nicht A0  ,  A1 und  nicht A13 und nicht A0  ,
ERROR, non high-live transition(s):
Boolean transition of type: CLOSING OR with arc-variables: A10, A11, A12
Boolean transition of type: CLOSING OR with arc-variables: A12, A13, A14
```

*Table 6: Model checking: EPC NOK*

After changing the logical type of transition $t_6$ into AND and of transition $t_7$ into XOR the resulting Boolean system $BS$ gets well-behaved, see Table 7. The Boolean system $BS$ has Boolean transitions of logical type AND and XOR only. Accordingly it can be replaced by its high system $BS^{high}$, which is the ordinary Petri net from Fig. 3 in [DVV2006] after short-circuiting.

```
Example 39: van der Aalst et al.: Verification of the SAP Reference Models ... 2006, corrected
OK, no deadlock
Successor graph: 1 strong component(s):
Minimal - Maximal - Component 1 with 3 node(s):
A1 und  nicht A13 und nicht A0  ,  nicht A1 und A13 und nicht A0  ,  nicht A1 und  nicht A13 und A0  ,
OK, high-live
```

*Table 7: Model checking: EPC OK*

# 6  Conclusion and Outlook

We start comparing our proposal for the model checking of Boolean process models with the results of the related papers named in the introduction from section 1, see Table 8.

| No. | Reference | Process model language | Reference language | Checked properties | Method of verification | Tool |
|---|---|---|---|---|---|---|
| 1. | [DVV2006] | EPC | ordinary safe Petri net | soundness, relaxed soundness | reduction | ProM |
| 2. | [MA2008] | EPC | EPC | EPC soundness | reduction, state space exploration | xoEPC |
| 3. | [FFJ2009] | UML/ BPMN | ordinary Petri net | safeness, liveness | model checking with CTL | LoLA |
| 4. | [FFJ2009] | UML/ BPMN | workflow net | soundness | reduction, state space exploration | Woflan |
| 5. | [FFJ2009] | UML/ BPMN | workflow graph | soundness | SESE fragmentation | IBM WebSphere Business Modeler |
| 6. | Present paper | EPC | Boolean system | safeness, liveness | model checking with propositional logic | under construction |

*Table 8: Selected methods for the verification of Boolean process models*

In our opinion the main differences between the methods from No. 1 to 5 compared to method 6 are the following:

- Method 6 considers a Boolean process model as a high-level construct and uses the high-level language of Boolean systems from the class of Coloured Petri nets. The other methods, which also use Petri nets as a reference language, always employ ordinary Petri nets.

  We think that the branching of the control flow as logical AND-, XOR-, OR-, AND_XOR- and other types of splits and joins cannot be adequately modelled by low-level constructs. Apparently each high-level Petri net can be flattened into an ordinary Petri net. But during this step much information gets lost which better should be kept together.

- Different from method 3, which is the only other model checking method from Table 8, method 6 uses model checking on high-level Petri nets. In our opinion high-level systems should be checked with high-level methods - as long as it is possible. For Boolean systems even the elementary means of propositional logic are sufficient.

- In our approach from [LSW1998] the semantics of an EPC is defined as the semantics of the corresponding Boolean system. Due to the concept of low-tokens the semantics is the usual Petri net semantics which is well-defined for each type of logical constructor.

  To the best of our knowledge we do not know other correct semantics of EPC constructs like the OR-join or the AND_XOR-join. We are well aware of different proposals in the literature, but often the proposed semantics is a global semantics and therefore seems at risk of the "vicious circle" [Kin2006].

- Tool support for method 6 is under construction. At present we are working on an interface between the tools CPN and Eclipse, in order to export Boolean system from CPN to Eclipse, the run-time environment of our implementation of Algorithm 4.9. We plan to link also a fully developed SAT-solver.

The theory of branching processes has been generalized from its original target of ordinary Petri nets to branching processes of high-level Petri nets [KK2003]. Different from this approach, which is based on low-level occurrence nets, our concept of a base process from Definition 4.1 studies a given high-level system by means of another high-level net, which is the colouring of a low-level occurrence net.

The purpose of our paper was to recall Boolean systems as a reference language for Boolean process models. The question, which of the advanced workflow patterns from [RHA2006] translate to the language of Boolean systems, has not been touched upon in the present paper: That's a challenge for further investigation.

Boolean systems are capable of providing a local semantics for the branching of the control flow according to arbitrary logical rules. Our method of verification applies to any strongly connected Boolean system. No further restrictions exist, neither with respect to the topology of the net nor concerning the initial marking, i.e., strongly connected Boolean systems with an arbitrary number of loops and initial tokens are admissible for model checking.